\documentclass[]{spie}  

\usepackage{amsmath,amsfonts,amssymb}
\usepackage{graphicx}
\usepackage[colorlinks=true, allcolors=blue]{hyperref}
\usepackage{siunitx}
\usepackage{chemformula}
\usepackage{subfigure}
\usepackage{multirow}
\usepackage{array}
\usepackage{wasysym}
\usepackage{xcolor}
\usepackage{arydshln}
\usepackage{diagbox}
\usepackage[table]{xcolor}
\usepackage{colortbl}

\title{Laser Remelting for Reduced Porosity on Additively Manufactured Aluminium Mirrors}

\author[a*,b]{Joshua West}
\author[a]{Valentina Oyarzun}
\author[a]{Marcell Westsik}
\author[a]{Younes Chahid}
\author[a]{Magdalena Kraus}
\author[a]{Scott McPhee}
\author[a]{William Brzozowski}
\author[c]{Sam Tammas-Williams}
\author[d]{Nicola Cayzer}
\author[e]{Fraser Laidlaw}
\author[f]{Sameer Dayanand Meshram}
\author[f]{Daniel Ordnung}
\author[f]{Berk Baris Celik}
\author[f]{Michel Smet}
\author[f,h]{Mirko Sinico}
\author[f]{Wenjuan Sun}
\author[f]{Brecht Van Hooreweder}
\author[g]{Michael Harris}
\author[g]{Stephen James}
\author[a**]{Carolyn Atkins}
\affil[a]{UK Astronomy Technology Centre, Royal Observatory, Edinburgh, EH9 3HJ, UK}
\affil[b]{Department of Mechanical Engineering, University of Bath, Bath, BA2 7AY, UK}
\affil[c]{School of Engineering, University of Edinburgh, Edinburgh, EH9 3JL, UK}
\affil[d]{School of Geoscience, University of Edinburgh, Edinburgh, EH8 9XP, UK}
\affil[e]{Physics and Astronomy, University of Edinburgh, Edinburgh, EH9 3FD, UK}
\affil[f]{Department of Mechanical Engineering, KU Leuven, 3001 Leuven, Belgium}
\affil[g]{Central Laser Facility Engineering and Technology Centre, STFC Rutherford Appleton Laboratory, Harwell, OX11 0QX, UK}
\affil[h]{Flanders Make@KU Leuven, 3001 Leuven, Belgium}

\authorinfo{*E-mail: jw3900@bath.ac.uk \\ **E-mail: carolyn.atkins@stfc.ac.uk}

\begin{document} 
\maketitle

\begin{abstract}
Additively manufactured (AM) AlSi10Mg mirrors are fabricated through laser powder bed fusion (LPBF), allowing the use of complex geometries such as lattices and organic structures that enable high mass reduction while maintaining mechanical stiffness. Micron-sized pores that cause optical scatter may form during LPBF as a consequence of deviations from the optimal processing window, particularly from laser energy input and scan strategy. This work proposes a laser remelting strategy aimed at reducing porosity; standard LPBF build steps automatically alternate with laser remelting passes, where previously deposited material is remelted during fabrication.

Laser remelting is evaluated through fabricating \SI{10}{\mm} proof-of-concept cubes. Following single point diamond turning (SPDT), optical measurements characterised surface roughness and identified surface artefacts. The best-performing AlSi10Mg remelted cube exhibited no pores within sampled regions and achieved \SI{6.4}{\nm} average surface roughness, comparable to a conventionally manufactured RSA 6061 control cube (\SI{5.8}{\nm}). Driven by these results, AM \diameter \SI{52}{\mm} secondary sandwich mirrors were manufactured using LPBF and laser remelting. These incorporate an optimised diamond TPMS lattice to achieve a 50\% mass reduction while accommodating design for AM considerations. Unlike the cube study, the optical surface of the remelted mirror after SPDT exhibited residual porosity and \SI{11.8}{\nm} average surface roughness. These results show that while the proof-of-concept confirmed the viability of laser remelting in reducing porosity within simple geometries, optimisation of the LPBF and SPDT processes are required to translate the benefits of laser remelting to lightweight AlSi10Mg AM mirrors.
\end{abstract}


\section{INTRODUCTION}
\label{sec:intro}

Aluminium mirrors for space applications must meet mass, volume, optical, and alignment requirements. Meeting these requirements simultaneously can be challenging with conventional manufacturing processes such as subtractive (milling) or formative (casting) due to limited tool access and the need for multiple fixtures and interfaces to perform these processes. Additive manufacturing (AM), which builds an object layer by layer from a 3D digital model, addresses this challenge. AM enables part consolidation, integrating multiple components into a single structure, thereby reducing part count and simplifying the design. Additionally, AM enables geometries which conventional manufacturing methods cannot, such as lattices and organic-style structures. When used within a mirror, these structures can provide mass reduction while maintaining mechanical stiffness.\cite{Zhang21} These capabilities have been demonstrated in previous projects such as \textit{Westsik et al. 2023}\cite{Westsik23}, where an AM mirror containing an internal lattice achieved a mass reduction of 44\% and consolidated nine parts into one. Within this context, the aluminium alloy AlSi10Mg has emerged as a common material choice for AM mirrors. AlSi10Mg is selected for its ease of processing in AM, commercial availability, and compatibility with single point diamond turning (SPDT), which is a post-processing technique often used to generate the optical surface. This compatibility is due to a fine microstructure, formed by rapid solidification of the metal alloy during the AM process\cite{tan24, Snell22, kotadia21}. Moreover, AlSi10Mg is commonly used in adjacent optomechanical structures which ensures a matching coefficient of thermal expansion (CTE) between the AM mirror and its assembly. Therefore, when under thermal loading in the extreme environments of space, all components distort with a similar expansion/contraction, reducing misalignments that could compromise the system\cite{chahid24}.

While AM achieves lightweight and compact aluminium mirror designs with ease, obtaining a high quality optical surface remains challenging. The optical performance of a mirror is driven by both the accuracy of the optical fabrication and material defects in the mirror. Surface form error represents the deviation of the surface shape from its optical prescription, contributing to wavefront error. Additionally, surface roughness characterises the texture of a surface and is ultimately limited by the microstructure/quality of the material. Surface roughness is characterised by the root mean square (RMS) or peak-to-valley (PV) of microscopic height differences on the surface, evaluated both over the entire optical surface and individual sampling regions. RMS provides a statistical measure of the average surface height variation, while PV identifies the maximum local height difference, allowing both the overall surface texture and isolated surface defects to be assessed. For shorter wavelengths, a smaller surface roughness is required. The success criteria of \SI{5}{\nm} RMS, as defined by \textit{Atkins et al. 2019}\cite{Atkins19-1}, for visible-wavelength performance was achieved overall on an AlSi10Mg AM mirror, with an estimated 2\% scatter. A surface roughness larger than the defined success criteria for a specific wavelength and/or application would cause excessive scatter, thus lowering image quality. Other studies that have fabricated AlSi10Mg AM mirrors have reported average surface roughness values from \SIrange{4.9}{5.8}{\nm} RMS after SPDT\cite{Aziz25, Lister24}, which also lie on or around the limit for the visible wavelength spectrum. However, the surface roughness measurements of individual sampling regions in these papers have reported major defects in the field of view (FOV), which have biased the overall RMS value. These major defects arise during the manufacturing process; unlike conventional manufacturing methods that begin with bulk material from cast or wrought stock, which is less likely to be porous or prone to contamination, AM processes can introduce defects within the material. In laser powder bed fusion (LPBF), the process used to fabricate AlSi10Mg AM mirrors, local melting and rapid solidification can promote pore formation owing to thermal effects. Depending on the process, additional contributors such as powder quality and raw material characteristics may also influence porosity. As a result, porosity remains a significant challenge for AM mirrors for space applications.

The LPBF process involves spreading a thin layer of AlSi10Mg powder, with a particle size distribution of D10 = \SI{23}{\um}, D50 = \SI{42}{\um}, and D90 = \SI{68}{\um}. Subsequently, this thin powder layer is selectively melted with a high-power laser. The melted powder forms the mirror geometry, and the unmelted powder supports subsequent layers. After one layer is complete, the build plate lowers and repeats the process. Variations in powder quality, LPBF parameters, or environmental conditions can trap gas within the melt pool, forming pores typically ranging from \SIrange{10}{100}{\um}\cite{Snell22, Liu22, Atkins18, Du23}. These pores have been observed both on the optical surface and within the mirror substrate in high frequencies\cite{Cooper19}. The occurrence of surface pores results in localised scatter which is reflected by increased surface roughness values in individual sampling regions. Defects and especially subsurface pores can drive crack growth during the lifetime of the component \cite{TammasWilliams2017}. These cracks can cause a weaker internal structure, especially if a lattice is used. Reducing porosity is therefore crucial to achieve a mirror that is lightweight, compact, and able to efficiently reflect infrared/visible wavelengths, and this paper focuses on methods to reduce this defect and improve optical surface quality.

\begin{figure}
    \centering
    \includegraphics[width=1\linewidth]{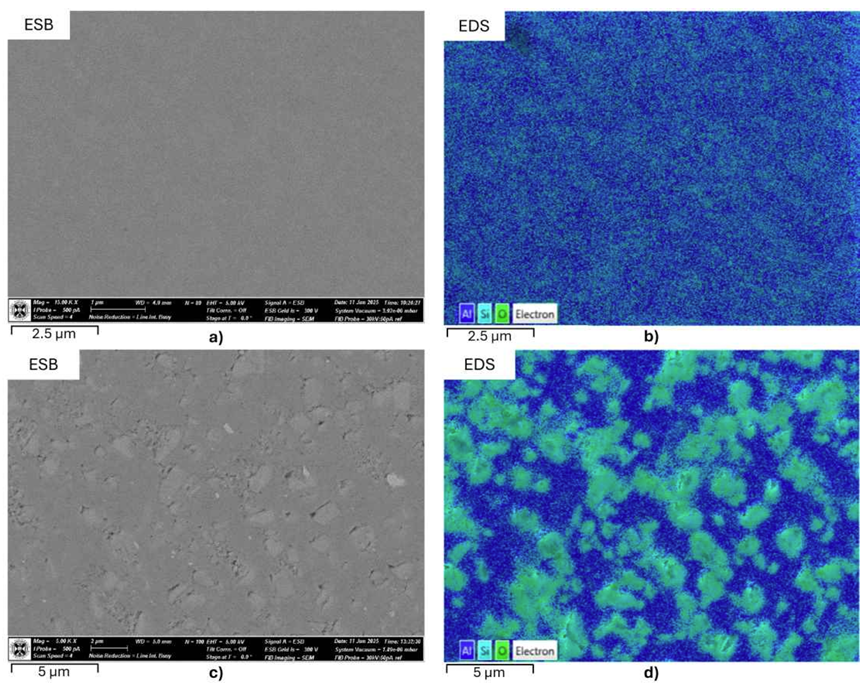}
    \caption{Reproduced from \textit{Aziz et al. (2025)}\cite{Aziz25}: High magnification scanning electron microscopy (SEM) and energy dispersive X-ray spectroscopy (EDX) images of mirrors with and without HIP; \textbf{a)} ESB image of non-HIP mirror optical surface; \textbf{b)} EDX scan of non-HIP mirror, same region as in \textit{(a)}; \textbf{c)} ESB image of HIP mirror optical surface; \textbf{d)} EDX scan of HIP mirror, same region as in \textit{(c)}.}
    \label{fig:aziz25silicon}
\end{figure}

Preventing the occurrence of pores is difficult due to an irregular distribution of thermal energy across layers. This is caused by the rapid melting and solidification of successive layers, combined with complex geometries such as lattices. Therefore, reducing porosity by optimising LPBF parameters is limited by mirror design\cite{Snell22, Lister24}. Post-processing techniques have been explored to reduce porosity in AM mirrors, including hot isostatic pressing (HIP). HIP is a post-processing technique done before SPDT to seal pores that have formed during the LPBF process before generating an optical surface\cite{Atkins2024, Aziz25}. HIP involves placing a component in a high heat and pressure environment for several hours, for AlSi10Mg this is at \SIrange{400}{550}{\degree}C and \SIrange{100}{300}{\MPa} for \SIrange{0.5}{2}{\hour}. Although this post-processing technique has proven to seal pores within the bulk material, a negative side effect also takes place. Within AlSi10Mg, the alloying silicon component has limited high temperature solubility, meaning that the heating of the mirror promotes silicon precipitation. Here, the silicon separates from the alloy and clusters\cite{kang25, sun24}. This produces heterogeneous regions in the material that causes protrusions to form on the surface (where the silicon has clustered) during SPDT, increasing surface roughness as seen in Figure \ref{fig:aziz25silicon}. A comparison between Figure \ref{fig:aziz25silicon} \textit{(b)} and \textit{(d)} highlights the conglomeration of silicon. These changes limit the improvement in optical surface quality, where the benefits of reduced porosity have been offset by the formation of silicon protrusions that increase surface roughness. Overall, as the improvement of optical quality is limited through optimising LPBF parameters and post-processing techniques, and an alternate approach to reducing porosity is required.

A laser remelting strategy, developed by KU Leuven, can be adapted to reduce porosity in AlSi10Mg AM mirrors. Laser remelting is a well-established strategy in LPBF that was originally developed to improve the quality of up-facing surfaces by reducing stress concentrations between melt pool boundaries and reducing the surface roughness of an as-built surface. In this work, laser remelting is adapted in a ProX DMP 320 machine from 3D Systems for pore reduction by alternating standard LPBF build steps with laser remelting passes using the build laser, so that previously deposited material is remelted during fabrication. The novelty of this work is the application of laser remelting to lightweight AM mirrors with complex lattice geometries to reduce porosity and evaluate its influence on the quality of an optical surface after SPDT post-processing.

This paper investigates the application of laser remelting on AM components fabricated from AlSi10Mg, and evaluates its effectiveness in reducing porosity to deliver high-quality optical surfaces after SPDT. Initially, a proof-of-concept study is conducted in Section \ref{sec:proofofconcept}, where \SI{10}{\mm} cube samples fabricated with varying laser remelting parameters investigate the effectiveness of laser remelting in reducing porosity at a simple geometric level. Driven by the findings of this proof-of-concept, Sections \ref{sec:specification} and \ref{sec:design} detail the specification and design of a first set of \diameter \SI{52}{\mm} lightweight sandwich mirrors for laser remelting. Internal lattices are used in the AM mirrors to achieve a 50\% mass reduction while preserving structural rigidity during manufacturing. Furthermore, a conventionally manufactured mirror is machined from an RSA 6061 billet and an AM mirror produced without laser remelting is made to act as a comparison. Section \ref{sec:manufacture} outlines the fabrication workflow and metrology procedure, in which the LPBF and laser remelting processes are followed by post-process subtractive machining and SPDT to generate the final optical surface for analysis. Finally, Section \ref{sec:results} provides a comparison between remelted mirrors, non-remelted mirrors, and conventionally machined RSA 6061 control substrates, employing various metrology methods to evaluate the quality of each optical surface, and the performance of laser remelting.

\section{LASER REMELTING PROOF-OF-CONCEPT}
\label{sec:proofofconcept}

A proof-of-concept study was conducted to evaluate the effectiveness of laser remelting on simple cube samples before applying the process to lightweight mirror geometries. This builds on an internal laser remelting parameter selection study performed by KU Leuven, where remelted cube samples fabricated in AlSi10Mg are repurposed for this proof-of-concept study to compare laser remelting against conventionally manufactured cube samples. With a goal of reducing porosity, this proof-of-concept study aims to produce and evaluate an optical surface on each cube to determine the effectiveness of laser remelting.

\subsection{Cube design and manufacturing}
\begin{figure}
    \centering
    \includegraphics[width=1\linewidth]{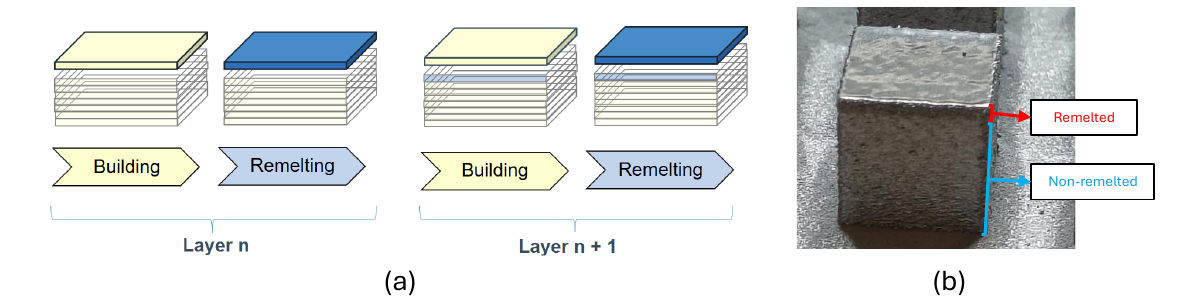}
    \caption{\textbf{a)} Schematic of the laser remelting process, where each standard LPBF build layer is immediately followed by a laser remelting pass before deposition of the next layer; \textbf{b)} Remelted cube on the build plate, with remelted and non-remelted regions annotated.}
    \label{fig:RemeltingDiagram}
\end{figure}

Nine cubes of \SI{10}{\mm} length were manufactured, named R2\_1 through R2\_9. During the manufacturing process, layers up to a height of \SI{9}{\mm} in all cubes were built without laser remelting, using LPBF parameters of \SI{350}{\W} laser power, \SI{1000}{\mm\//\s} scan speed, \SI{100}{\um} hatch spacing, and a \SI{30}{\um} height step. The remaining \SI{1}{\mm} of layers underwent laser remelting, where LPBF building steps alternated with laser remelting steps as seen in Figure \ref{fig:RemeltingDiagram}. The laser remelting parameters used are listed in Table \ref{tab:CubePrintingParameters}, with R2\_3 made without laser remelting (non-remelted). After LPBF and laser remelting, the cubes were removed from the build plate via wire electric discharge machining (EDM). Bulk porosity measurements in the internal KU Leuven study identified R2\_9 as the cube with the least porosity, therefore, its parameters were selected for the first set of mirrors.

\begin{figure}
    \centering
    \includegraphics[width=1\linewidth]{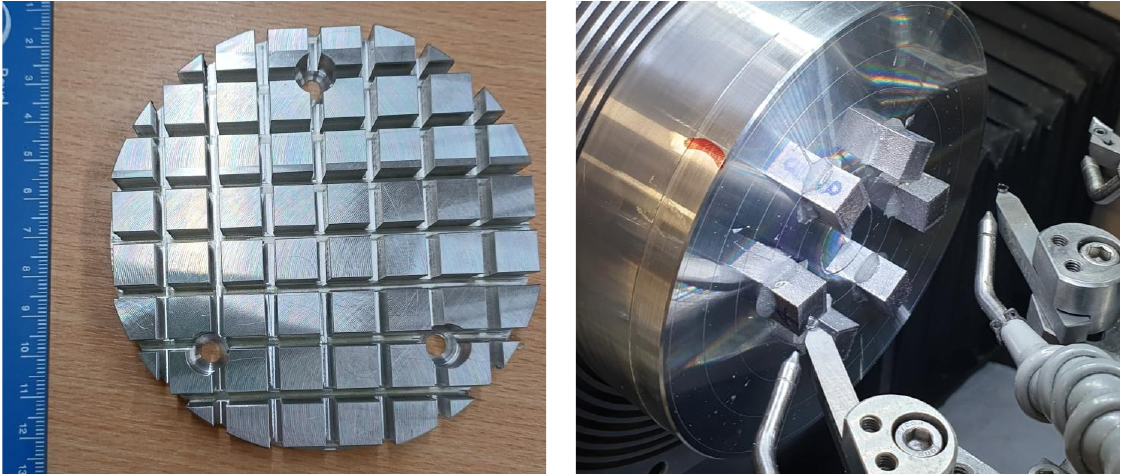}
    \caption{\textit{left} - the control cubes machined into a RSA 6061 cast billet; \textit{right} - the single point diamond turning arrangement to generate an optical surface on the cube samples. All cubes are turned off-axis, clockwise. Image credit M. Harris.}
    \label{fig:CubeSPDT}
\end{figure}

Three control cubes were manufactured from RSA 6061, an aluminium alloy that is frequently favoured for mirror fabrication with SPDT\cite{Guido08, Katgerman04}. This manufacturing process consisted of machining \SI{10}{\mm} length cubes from a \diameter \SI{105}{\mm} RSA 6061 billet, shown in Figure \ref{fig:CubeSPDT} \textit{(a)}, then removing the cubes via wire EDM. These cubes were named RSA\_1 through RSA\_3.

All cubes then underwent SPDT, as displayed in Figure \ref{fig:CubeSPDT} \textit{(b)}, to generate an optical surface on each cube. The setup of the SPDT process is based on in-house expertise, with the goal of generating a surface with least surface roughness. To minimise scratches, material is initially removed using \SI{0.050}{\mm} cut depths before progressively reducing the depth of cut to minimise forces applied at the cutting tip of the diamond tool on the final passes.

{\renewcommand{\arraystretch}{1.25}
\begin{table}[h]
\centering
\caption{Laser remelting parameters investigated in the proof-of-concept study, R2\_3 is non-remelted.}
\newcolumntype{C}[1]{>{\centering\arraybackslash}m{#1}}
\begin{tabular}{|C{1.2cm}||C{1.3cm}|C{2.5cm}|C{1.4cm}|C{2.0cm}|C{1.2cm}|C{2.8cm}|}
\hline
\textbf{Cube} &
\textbf{Power {[}W{]}} &
\textbf{Scan Speed {[}mm/s{]}} &
\textbf{Hatch {[}$\pmb{\mu}$m{]}} &
\textbf{Height Step {[}$\pmb{\mu}$m{]}} &
\textbf{Passes} &
\textbf{Scan Rotation {[}deg{]}} \\
\hline
R2\_1 & 350 & 700 & 70 & 30 & 2 & 90 \\
R2\_2 & 350 & 700 & 70 & 30 & 4 & 90 \\
R2\_3 & -- & -- & -- & -- & -- & -- \\
R2\_4 & 250 & 700 & 70 & 30 & 2 & 90 \\
R2\_5 & 250 & 700 & 70 & 30 & 4 & 90 \\
R2\_6 & 200 & 700 & 70 & 30 & 4 & 90 \\
R2\_7 & 200 & 700 & 70 & 30 & 2 & 90 \\
R2\_8 & 250 & 700 & 35 & 30 & 2 & 90 \\
R2\_9 & 350 & 700 & 70 & 30 & 2 & 67 \\
\hline
\end{tabular}
\label{tab:CubePrintingParameters}
\end{table}}

\subsection{Cube evaluation}

AlSi10Mg components produced via LPBF can exhibit a range of surface and sub-surface artefacts. In this study, the primary artefacts of interest are pores, oxides, and scratches found on the optical surface. Oxides form during the LPBF process, where oxygen not removed from the build chamber by the inert gas can be trapped within the aluminium powder. These oxides are harder than the fused aluminium and have been shown to initiate scratches during the SPDT process.\cite{Bai20, Atkins26} Scratches can be categorised based on whether they are caused by underlying issues from the LPBF process and laser remelting (e.g., oxides and contaminants embedded in the optical surface). Each can be identified based on their alignment with the SPDT tool path, origin from a visible scratch initiator on the surface, and same direction of travel to the cutting direction of the SPDT tool. Scratches caused by external factors that are outside this study are those that do not comply with the criteria above, including:
\begin{itemize}
    \item Scratches initiated from the edge of the cube - the chamfers of the cube were known to be rough, and material was easily picked up by the SPDT tool causing a scratch.
    \item A scratch not aligned with the SPDT tool path - this is induced from wiping the surface clean after SPDT.
\end{itemize}

To identify these artefacts and thoroughly examine the cubes' surfaces, a specific set of analysis methods must be applied: optical microscopy, scanning electron microscopy (SEM), energy dispersive X-ray spectroscopy (EDX), focused ion beam (FIB), and measurement of surface roughness. Combined, these processes can evaluate the quality of an optical surface through identifying artefacts and quantifying the overall characteristics of an optical surface.

For optical microscopy, an Evident DSX2000 with a focal length of \SI{35}{\mm} is used to generate stitched composite bright-field (BF) and dark-field (DF) images of the optical surfaces at a $ \times $10 magnification. This optical microscopy setup measures the cube surface using reflected light and has a limited imaging resolution compared to alternative analysis methods. These attributes allow a global sampling of the \SI{10}{\mm} $ \times $ \SI{10}{\mm} surface while highlighting micron-sized surface artefacts, but provides a restricted understanding relating to artefact morphology and composition. This is a limitation of the analysis method, and further examination of the optical surface is required to identify what certain surface artefacts are. All cubes were measured with optical microscopy.

A Zeiss Crossbeam 550 FIB-SEM was used to perform SEM, EDX, and FIB-SEM imaging of surface defects. SEM has the capability of identifying the morphology and limited compositional information through the backscatter electrons of a surface artefact and EDX provides further chemical composition information of a surface artefact. FIB can create micron-sized cuts into the surface of a component, which can extend the 2D morphology of a surface artefact into a 3D cross-section allowing further evaluation of surface artefact morphology and composition by SEM and EDX. Overall, the Crossbeam uses a limited FOV and image capture is time consuming; therefore, the optical microscopy images are first used to identify areas of interest before being evaluated by the Crossbeam, enabling efficient targeted analysis at higher magnification. Only cubes R2\_9 and RSA\_3 were analysed with SEM, EDX, and FIB, as these cubes were the representative remelted and control samples identified in microscopy in Section \ref{subsec:cuberesults}.

\begin{figure}
    \centering
    \includegraphics[width=1\linewidth]{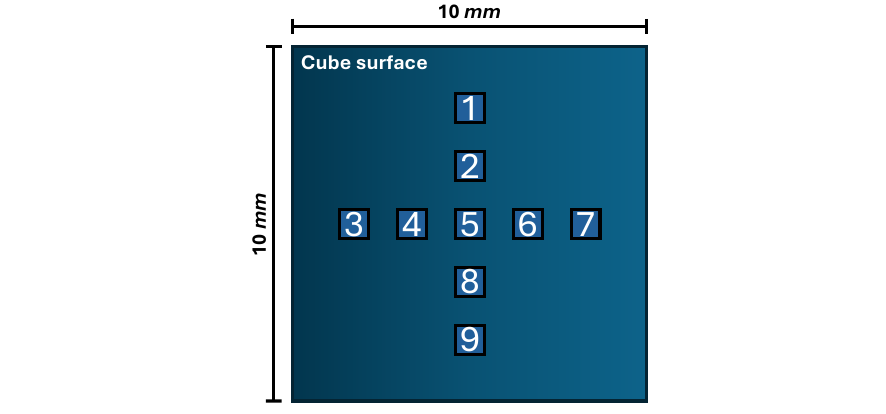}
    \caption{Layout of surface roughness sampling arrangement for cube surface.}
    \label{fig:cubelayout}
\end{figure}

Surface roughness measurements are obtained using a Bruker Contour-X 200 with a $ \times $20 magnification objective in phase shifting interferometry (PSI) mode, with an FOV comparable to SEM (at specific SEM magnifications) at \SI{410}{\um} $ \times $ \SI{344}{\um}. The Bruker uses an automated $ x $-$ y $ stage and the cubes were sampled in an inner square of \SI{6}{\mm} $ \times $ \SI{6}{\mm} segmented into nine \SI{1.5}{\mm} $ \times $ \SI{1.5}{\mm} measurements in a vertical-cross ($+$) arrangement, as depicted in Figure \ref{fig:cubelayout}. A Gaussian regression filter with a cut-off of $ \lambda_c = $ \SI{80}{\um} was applied in accordance with ISO 16610 and ISO 25178 to remove form error from the roughness data, providing the mean (Sa), RMS (Sq), and PV (Sz) of the surface roughness of each segment. Only the surface roughness of cubes R2\_9 and RSA\_3 were measured as these cubes were the representative remelted and control samples identified in microscopy in Section \ref{subsec:cuberesults}.

\subsection{Cube results}
\label{subsec:cuberesults}

\begin{figure}
    \centering
    \includegraphics[width=1.0\linewidth]{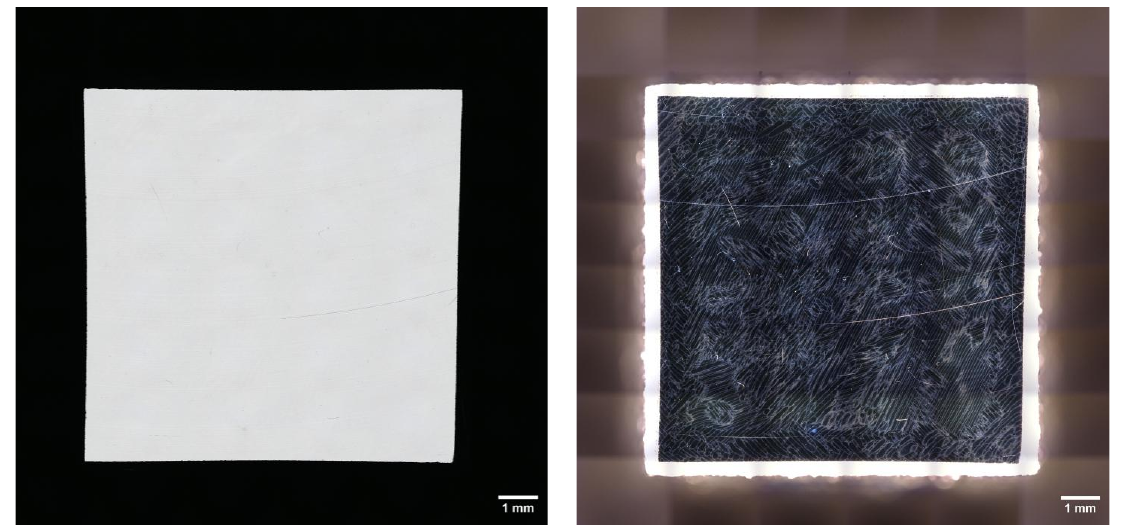}
    \caption{Microscopy images of R2\_9: \textit{left} - bright-field, at a global perspective only deep cuts visible; \textit{right} - dark-field, melt pool boundaries and SPDT cutting tracks visible.}
    \label{fig:R2_9_Microscopy}
\end{figure}

\begin{figure}
    \centering
    \includegraphics[width=0.8\linewidth]{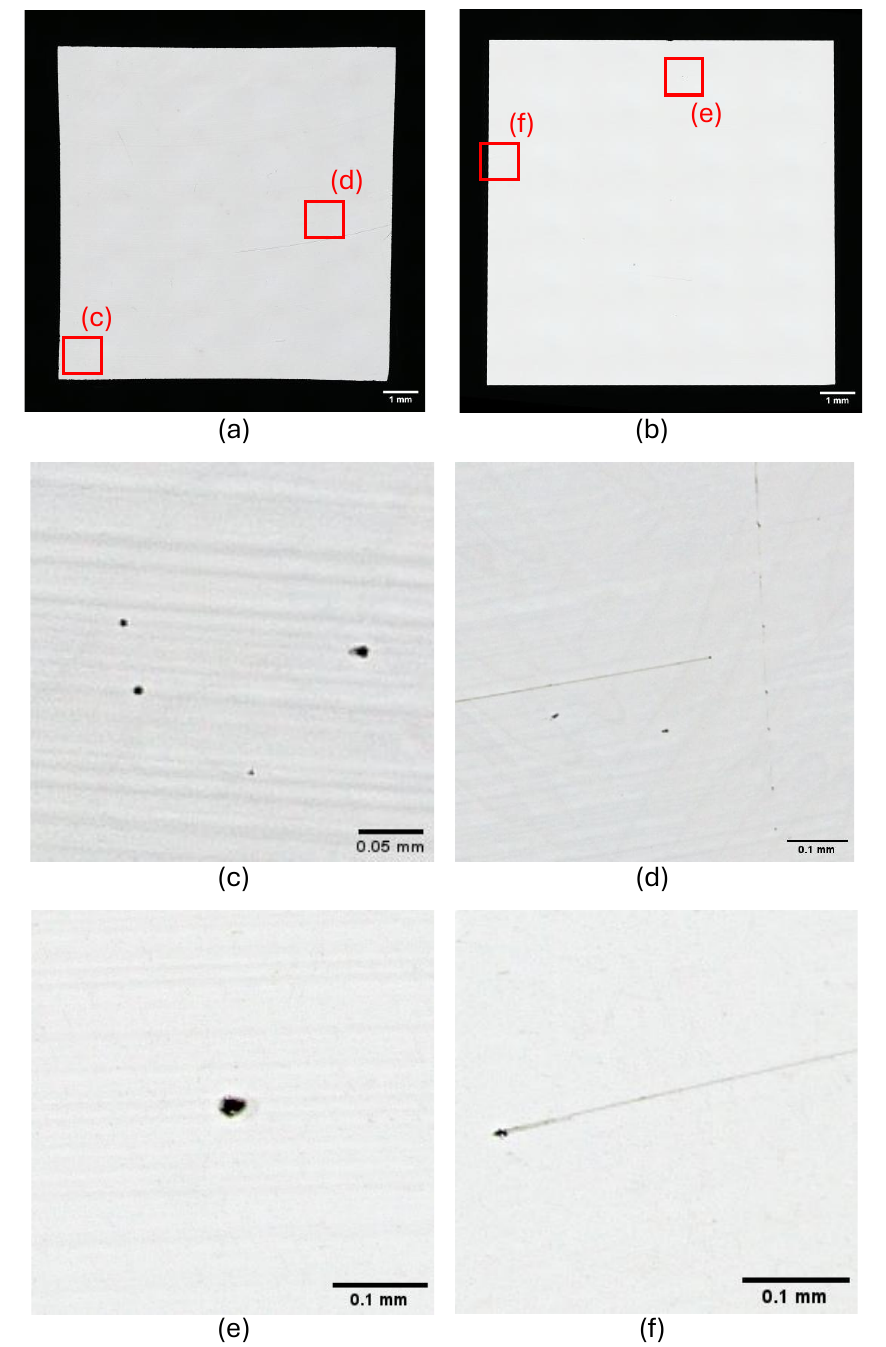}
    \caption{Bright-field microscopy images: \textbf{a)} R2\_9 whole optical surface; \textbf{b)} RSA\_3 whole optical surface; \textbf{c)} detail of R2\_9 \textit{(a)}, black spots - likely pore or oxide; \textbf{d)} detail of R2\_9 \textit{(a)}, scratch initiated by SPDT on surface; \textbf{e)} detail of RSA\_3 \textit{(b)}, black spot - pore or oxide; \textbf{f)} detail of RSA\_3 \textit{(b)}, scratch initiated by SPDT on surface.}
    \label{fig:R2_9_Defects}
\end{figure}

Optical microscopy identified surface artefacts on the optical surface of all cubes. Figure \ref{fig:R2_9_Microscopy} shows the BF and DF images of R2\_9, which are used to highlight different surface artefacts:

\begin{itemize}
    \item \textbf{Primary - Pores \& Oxides:} Pores and oxides both appear as spherical and irregular-shaped black spots due to low reflectivity on the surface of both the remelted and control cubes as observed in Figure \ref{fig:R2_9_Defects} \textit{(c)} and \textit{(e)}, meaning their true form (i.e., whether the spot is a pore or oxide) cannot be confirmed from microscopy alone. For spherical or near-spherical black spots, the equivalent diameter can be measured. This value is measured through the software ImageJ by taking the area of a black spot and calculating the diameter if assumed circular and a pore. As this value follows assumptions, it can only give a vague idea of pore diameter and a conclusion cannot be drawn. The average equivalent diameter of sampled black spots from the BF image of R2\_9 was \SI{9.26}{\um}, a value smaller than common sampling thresholds for pore analysis, suggesting the black spots are not pores.\cite{Chahid24-pore}
    \item \textbf{Secondary - Scratches:} Scratches caused by underlying issues from the LPBF process and laser remelting are seen in sub Figures \ref{fig:R2_9_Defects} \textit{(d)} and \textit{(f)}.
\end{itemize}

Among the remelted cubes, R2\_9 had the least black spots (pores or oxides) visible on the optical surface, which supports the findings of the internal KU Leuven study that this sample had low porosity. All RSA control cubes presented black spots \& scratches similar to those seen on the remelted cubes, highlighting the optical surfaces were not perfect. As both remelted and control cubes displayed similar surface artefacts under optical microscopy, R2\_9 and RSA\_3 were taken forward for further SEM, EDX, and FIB analysis and surface roughness measurement.

\begin{figure}
    \centering
    \includegraphics[width=0.8\linewidth]{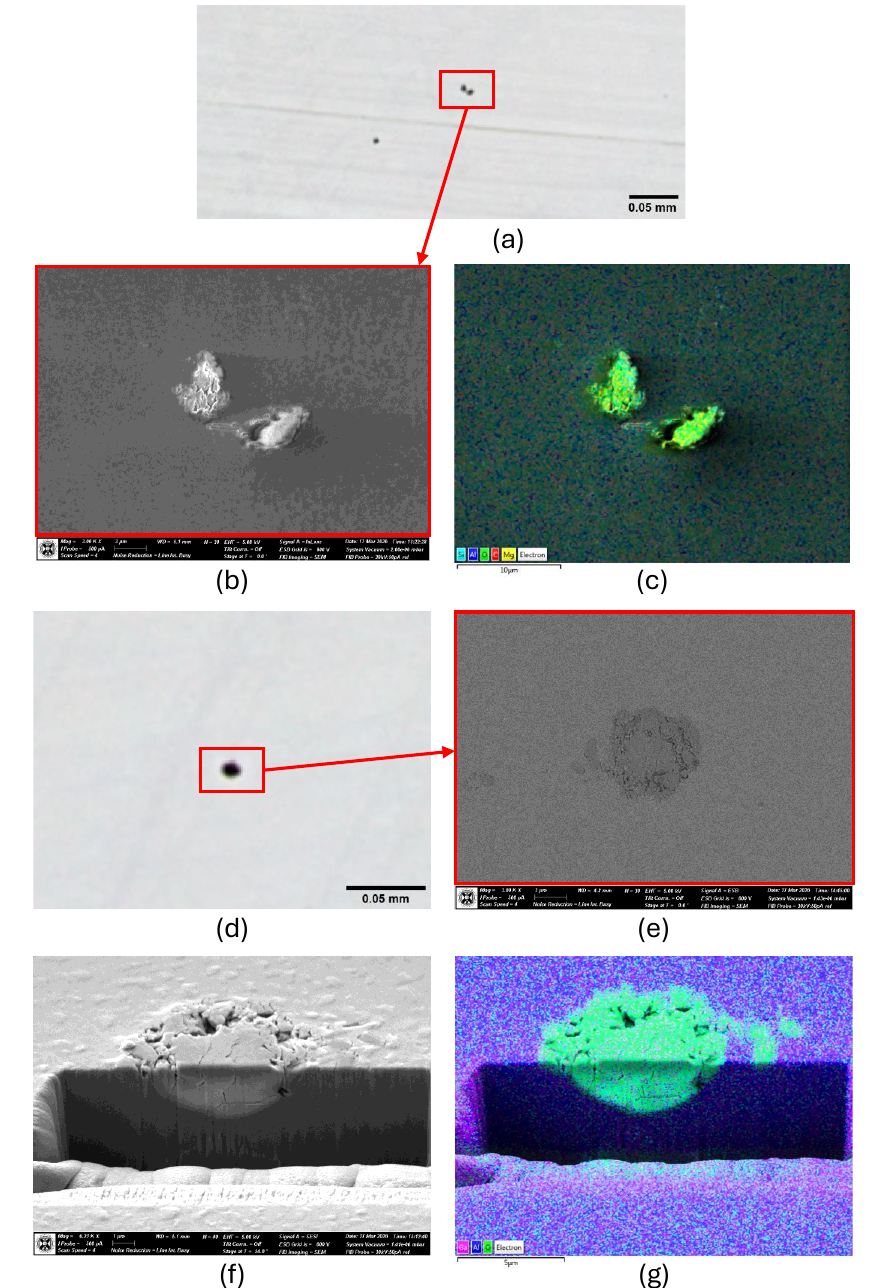}
    \caption{Microscopy and SEM characterisation of two representative surface artefacts on R2\_9. Artefact 1: \textbf{a)} bright-field microscopy image showing a black spot; \textbf{b)} SEM image showing detailed view; \textbf{c)} EDX elemental map overlay confirming surface artefact is an oxide. Artefact 2: \textbf{d)} bright-field microscopy image of a black spot; \textbf{e)} SEM image showing detailed view; \textbf{f)} focused ion beam (FIB) cross-section through surface artefact showing subsurface morphology; \textbf{g)} EDX elemental map overlay confirming surface artefact is an oxide.}
    \label{fig:R2_9_SEM1}
\end{figure}

\begin{figure}
    \centering
    \includegraphics[width=1.0\linewidth]{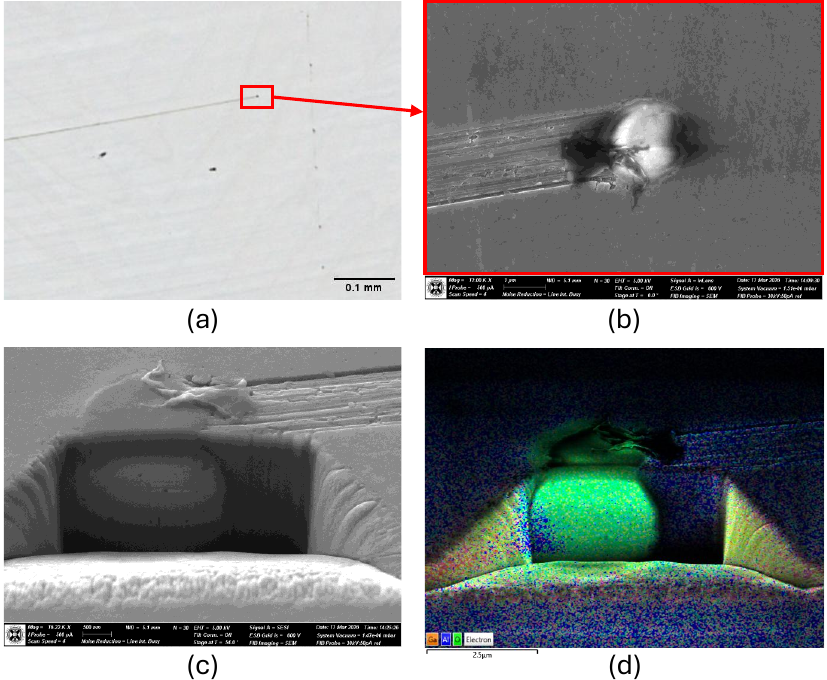}
    \caption{Microscopy and SEM characterisation of an oxide-initiated scratch on R2\_9: \textbf{a)} bright-field microscopy image showing scratch location; \textbf{b)} SEM image showing detailed view of scratch initiator; \textbf{c)} focused ion beam (FIB) cross-section through scratch initiator, revealing subsurface morphology; \textbf{d)} EDX elemental map overlay, shows larger oxide beneath is larger relative to that seen on the surface (smaller cross-section on surface).}
    \label{fig:R2_9_SEM2}
\end{figure}

Images using SEM, EDX, and FIB on the black spots of R2\_9 and RSA\_3 confirmed that they were all oxides. All oxide regions had similar elemental composition throughout both cubes: aluminium and oxygen with minor contributions ($<5$ weight \%) of silicon, 
 and carbon, as seen in Figure \ref{fig:R2_9_SEM1} which shows two cases of the oxide on R2\_9. For comparison, a surface artefact known to be an oxide (due to initiating a scratch) is seen in Figure \ref{fig:R2_9_SEM2}, and has a similar composition to the oxides seen in Figure \ref{fig:R2_9_SEM1}. The oxide in Figure \ref{fig:R2_9_SEM2} underwent FIB, revealing the subsurface morphology of the oxide. The FIB cross-section shows that the oxide extends below the optical surface, indicating that it is embedded within the material rather than being only a surface-level artefact. Table \ref{tab:SurfaceDefectCounts} summarises the number of each defect type sampled during SEM, EDX, and FIB of the cubes. Overall, no pores were identified on the optical surfaces of either R2\_9 or RSA\_3 within the sampled regions.

{\renewcommand{\arraystretch}{1.25}
\begin{table}[h]
\centering
\caption{Number of surface defects sampled by type, on R2\_9 and RSA\_3 samples.}
\newcolumntype{C}[1]{>{\centering\arraybackslash}m{#1}}
\begin{tabular}{|m{4cm}||C{2cm}|C{2cm}|}
\hline
\diagbox[width=4.45cm]{\ \ \textbf{Defect Count}}{\textbf{Sample}\ \ \ } &
\textbf{R2\_9} & \textbf{RSA\_3} \\
\hline
Oxides & 7 & 2 \\
Pores & 0 & 0 \\
Scratches & 1 & 2 \\
Contaminant / Organic & 2 & 2 \\
\hline
\end{tabular}
\label{tab:SurfaceDefectCounts}
\end{table}}

\begin{figure}
    \centering
    \includegraphics[width=1\linewidth]{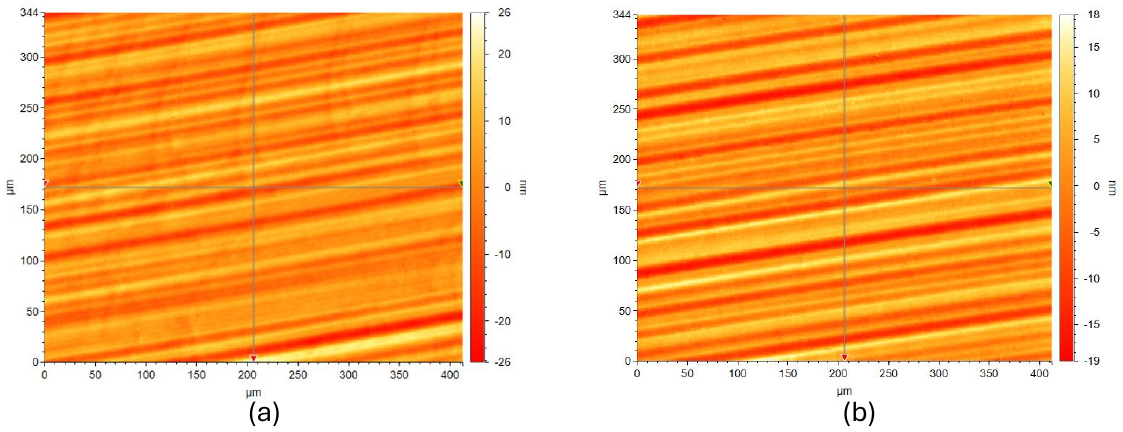}
    \caption{Surface roughness measurement of R2\_9 and RSA\_3: \textit{a)} R2\_9; \textit{b)} RSA\_3. The only difference is that the melt pools of R2\_9 are slightly visible; however, the cutting passes remain dominant.\newline}
    \label{fig:cubesroughness}
\end{figure}

The surface roughness values of R2\_9 and RSA\_3 were measured. As seen in Figure \ref{fig:cubesroughness}, the surface roughness of both cubes is dominated by the cutting passes of SPDT. Note that in Figure \ref{fig:cubesroughness} \textit{(a)} the melt pool boundaries of R2\_9 slightly influences the surface roughness, whereas in Figure \ref{fig:cubesroughness} \textit{(b)} as RSA\_3 is conventionally manufactured it is only influenced by the cutting marks.

{\renewcommand{\arraystretch}{1.25}
\begin{table}[h]
\centering
\caption{Surface roughness measurements for R2\_9 and RSA\_3 to 1 decimal place, with calculated mean, min, max, and standard deviation for each. The high PV (Sz) values are due to surface artefacts within the sampled region.} 
\newcolumntype{C}[1]{>{\centering\arraybackslash}m{#1}}
\begin{tabular}{|m{3cm}||C{1.4cm}|C{1.4cm}||C{1.4cm}|C{1.4cm}||C{1.4cm}|C{1.4cm}|}
\hline
& \multicolumn{2}{c||}{\textbf{Mean (Sa) {[}nm{]}}} & \multicolumn{2}{c||}{\textbf{RMS (Sq) {[}nm{]}}} & \multicolumn{2}{c|}{\textbf{PV (Sz) {[}nm{]}}} \\  
\hline
\diagbox[width=3.445cm]{\ \ \ \textbf{Data}}{\textbf{Cube}\ \ \ }& \textbf{R2\_9} & \textbf{RSA\_3} & \textbf{R2\_9} & \textbf{RSA\_3} & \textbf{R2\_9} & \textbf{RSA\_3} \\
\hline
Position 1 & 4.8 & 4.6 & 6.1 & 5.8 & 57.1 & 36.0 \\
Position 2 & 4.5 & 3.8 & 5.9 & 5.0 & 391.0 & 37.0 \\
Position 3 & 5.4 & 5.0 & 6.8 & 6.2 & 50.1 & 37.3 \\
Position 4 & 4.7 & 4.4 & 5.9 & 5.6 & 44.9 & 36.3 \\
Position 5 & 5.2 & 4.5 & 6.6 & 5.8 & 93.5 & 40.1 \\
Position 6 & 5.2 & 4.4 & 6.7 & 5.7 & 52.4 & 37.0 \\
Position 7 & 5.1 & 4.2 & 6.5 & 5.3 & 58.1 & 72.0 \\
Position 8 & 5.0 & 5.5 & 6.4 & 7.0 & 127.0 & 45.9 \\
Position 9 & 5.1 & 4.8 & 6.4 & 6.2 & 61.1 & 48.3 \\
\hline
Mean & 5.0 & 4.6 & 6.4 & 5.8 & 103.9 & 43.3 \\
\cdashline{1-7}
Min & 4.5 & 3.8 & 5.9 & 5.0 & 44.9 & 36.0 \\
\cdashline{1-7}
Max & 5.4 & 5.5 & 6.8 & 7.0 & 391.0 & 72.0 \\
\cdashline{1-7}
Standard Deviation & 0.3 & 0.5 & 0.3 & 0.6 & 110.8 & 11.7 \\
\hline
\end{tabular} 
\label{tab:CubeSurfaceRoughnessData}
\end{table}}

\subsection{Proof-of-concept conclusion}

This proof-of-concept study evaluated the suitability of laser remelting for AM mirrors and its ability to reduce porosity on an optical surface by comparing nine remelted AlSi10Mg cubes to conventionally manufactured RSA 6061 control cubes. Optical microscopy revealed surface artefacts across all samples, with black spots initially interpreted as pores or oxides. The remelted cube R2\_9 and control cube RSA\_3 were taken forward for further analysis. SEM, EDX, and FIB confirmed that the sampled black spots on both cubes were oxides rather than pores. Notably, no pores were identified in any sampled region of R2\_9.

The surface roughness measurements indicate that SPDT cutting marks are the dominant contributor for both R2\_9 and RSA\_3, as seen in the defined surface texture in both cases. Although the average RMS (Sq) surface roughness of R2\_9 is \SI{0.6}{\nm} larger than RSA\_3, the observed difference is small compared to the standard deviation of the measurements, indicating no clear separation in surface roughness between the two cubes. Nevertheless, if the surface roughness values of both cubes qualifies for a target wavelength, the advantage of having a \SI{0.6}{\nm} improvement in surface roughness for RSA\_3 can be outweighed by the advantages enabled by AM for R2\_9 through having unconstrained geometry, lightweighting through internal structures, and consolidated features.

Overall, the results indicate that with optimised parameters, laser remelting can suppress pores within the LPBF produced bulk material, and allow the production of an optical surface via SPDT with a surface roughness comparable to conventionally manufactured aluminium mirrors. This supports the potential of the laser remelting strategy and its application in this work, where it is a viable process route for AlSi10Mg AM mirrors for porosity mitigation.

\section{MIRROR DESIGN SPECIFICATION AND OBJECTIVES}
\label{sec:specification}

Based on the proof-of-concept, the focus of this project is to apply laser remelting to AlSi10Mg AM mirrors and evaluate its effectiveness in reducing porosity, as seen with the proof-of-concept cubes. This requires a design specification that allows the mirror design and manufacture to incorporate laser remelting while maintaining the core aspects of a functional lightweight mirror by accommodating both a lightweight lattice and an optical surface.

\subsection{Design for additive manufacturing considerations and constraints}

To generate a design specification for this application, considerations specific to AM must be made. Although AM provides geometric freedoms that exceed conventional manufacturing methods in terms of geometry complexity, certain constraints must be followed to apply the benefits of AM, and ensure the successful fabrication of a component:

\begin{itemize}
    \item \textbf{Geometric}: General \textit{DfAM rules}\cite{Atkins22} apply to any AM component, including minimum feature size, required build resolution, overhang limits, supported and unsupported geometries, and build orientation. However, additional considerations are required for components made via LPBF, such as designing a continuous conductive path within the component to allow heat to dissipate. This is especially important for AM mirrors that contain complex structures such as lattices, where unfused powder acts as an insulator. Therefore, the LPBF parameters optimised for manufacturing solid material (the optical surface) do not translate to the parameters used in the lattice, meaning the chosen lattice type and parameters need to accommodate both DfAM rules and LPBF considerations within the lattice.
    \item \textbf{Post-processing}: The AM process alone cannot achieve the optical performance and optomechanical mounting requirements of a mirror. Post-processing is needed to generate the final optical surface and precise features such as the mounting holes. So, the mirror must be designed around enabling this, where it has an accessible geometry and a sufficient stiffness to withstand machining and optical fabrication processes.
    \item \textbf{Powder removal}: During LPBF, solid material is built around unmelted powder; depending on the geometry, this powder can become entrapped within the part. Entrapped powder is undesirable as it increases mass and alters the centre-of-mass, which can interfere with post-processing, expected performance, and poses a health \& safety risk. Therefore, the design of the mirror must incorporate a path for powder exit, allowing the complete removal of residual powder. 
\end{itemize}

\subsection{Lightweight mirror design}

\begin{figure}
    \centering
    \includegraphics[width=0.95\linewidth]{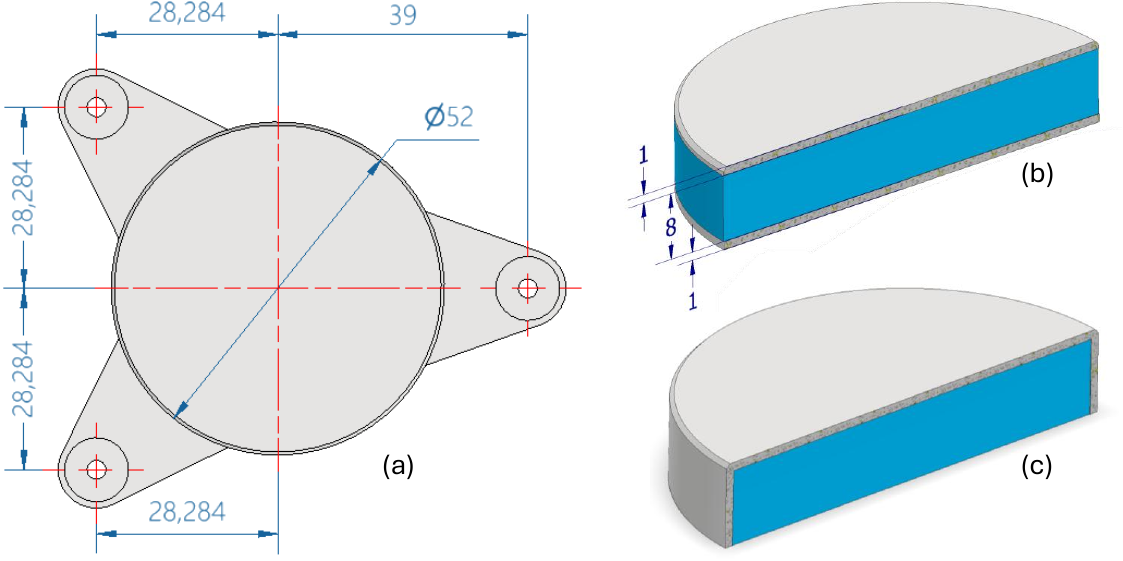}
    \caption{Mirror geometries, the lattice volume is represented by the blue solid. \textit{(b)} and \textit{(c)} are half-section representations to highlight the lattice form: \textbf{a)} on-axis view of mirror, surface and mount geometries measured; \textbf{b)} Sandwich mirror design -- the lattice is sandwiched by base and face plates, and exposed through the side walls; \textbf{c)} open-back mirror design -- the lattice is enclosed by solid side walls and exposed through the base plate.}
    \label{fig:MirrorDesigns}
\end{figure}

Considering the novelty of the laser remelting process to mirror fabrication and the in-house AM mirror heritage, a secondary mirror (M2) from an in-house nano-satellite application is chosen, which has a simple geometry (\diameter \SI{52}{\mm} and \SI{10}{\mm} height). This mirror geometry allows the design, manufacture, and testing of the mirror to build upon and advance an earlier study by \textit{Lister et al. (2024)}\cite{Lister24}, since the mirror has the same geometry. Furthermore, to allow for post-processing, a SPDT fixture used in the earlier study is reused in this project to reduce waste. Therefore, three mounting holes are integrated into the lightweight mirror design at known locations, seen in Figure \ref{fig:MirrorDesigns} \textit{(a)}.

Lightweight mirror designs can be characterised by which face the lightweight structure is exposed in relation to the optical surface. Sandwich and open-back designs are common lightweight mirror designs, described in Figure \ref{fig:MirrorDesigns} \textit{(b)} and \textit{(c)}. To allow for powder removal, an open-back design is typically manufactured at an inclined orientation using removable supports. This inclination also avoids a large horizontal overhang of the optical surface, becoming self-supporting.\cite{Tan20, Zhang22} In contrast, a sandwich mirror design can be manufactured with the optical surface parallel to the build plate, where the exposed side walls provide access to the lattice for powder removal, removing the need for supports. However, this orientation removes the self-supporting benefit of an inclined build, where the optical surface is supported solely by the underlying lattice during manufacture. In this paper, laser remelting requires the optical surface to remain parallel to the build plate during fabrication to maximise the effectiveness of laser remelting\cite{Ordnung24}. Furthermore, manufacturing an optical surface in-plane with the build plate benefits the resulting microstructure, as the grains tend to grow normal to the build plate. Therefore, a sandwich mirror was selected to facilitate the laser remelting process and to grow in-house mirror design expertise. With this information, Table \ref{tab:DesSpec} summarises a design specification that defines all criteria that must be met for a valid lightweight mirror design for laser remelting.

{\renewcommand{\arraystretch}{1.25}
\begin{table}[h]
\centering
\caption{Mirror design specification, separated by parameter categories: geometric, optical, design for additive manufacturing (DfAM), and post-processing.}
\begin{tabular}{|l||l|l|}
\hline
\textbf{Type} & \textbf{Parameter} & \textbf{Specification} \\
\hline
\multirow{12}{*}{Geometric}
& Substrate Architecture & Cylindrical sandwich mirror \\
& Mechanical Aperture & \diameter \SI{52}{\mm} \\
& Optical Aperture & \diameter \SI{50}{\mm} \\
\cdashline{2-3}
& Total Substrate Height & \SI{10}{\mm} \\
& \quad Face Plate & \SI{1}{\mm} \\
& \quad Lattice Design Space & \SI{8}{\mm} \\
& \quad Base Plate & \SI{1}{\mm} \\
\cdashline{2-3}
& Mounting / Alignment Features & Clearance hole, 3 $ \times $ \diameter \SI{3.1}{\mm} \\
& \quad Origin & Base plate, face plane, on axis \\
& \quad Position 1 & (39, 0) \SI{}{\mm} \\
& \quad Position 2 & (-28.284, 28.284) \SI{}{\mm} \\
& \quad Position 3 & (-28.284, -28.284) \SI{}{\mm} \\
\hline
\multirow{2}{*}{Optical}
& Optical Prescription & Flat \\
& Radius of Curvature (ROC) & ------ \\
\hline
\multirow{6}{*}{DfAM}
& Material & AlSi10Mg \\
& Print Direction & Parallel to build plate, face plate upwards\\
& Minimum Wall Thickness & \SI{0.5}{\mm} \\
& Powder Removal & All residual powder removed \\
& Overhangs & No large overhangs over $ 45^\circ $ \\
& Supports & Minimal; as required \\
\hline
\multirow{3}{*}{Post-processing}
& Additional Face Plate Material & \SI{3}{\mm} \\
& Additional Base Plate Material & \SI{1}{\mm} \\
& Optical Surface Feature & \SI{0.5}{\mm} $ \times $ $ 45^\circ $ Chamfer \\
\hline
\end{tabular}
\label{tab:DesSpec}
\end{table}}

\subsection{Design Objectives}
\label{subsec:desobjs}

The following design objectives will be used to generate a mirror ideal for laser remelting. Where the design specification in Table \ref{tab:DesSpec} must also be met.

\begin{enumerate}
    \item Achieve a target mass reduction of 50\% for the mirror (excluding mounting legs) when compared to a solid equivalent; mounting legs are excluded as only hole locations are defined. Nevertheless, topology optimisation (TO) will be used to minimise mass of the legs while maintaining sufficient stiffness for post-processing.
    \label{itm:desobj1}
    \item The mirror shall be compatible with both standard LPBF and laser remelting.
    \label{itm:desobj2}
    \item Surface displacement during SPDT shall remain uniform across the optical surface.
    \label{itm:desobj3}
\end{enumerate}

\section{MIRROR DESIGN}
\label{sec:design}

With the mirror design specification and objectives defined, the design process involves implementing and optimising the lightweight lattice within the predefined lattice region, while generating mounting features to allow post-processing operations after fabrication.

\subsection{Lattice Selection}
\label{subsec:latticeselection}

\begin{figure}
    \centering
    \includegraphics[width=1\linewidth]{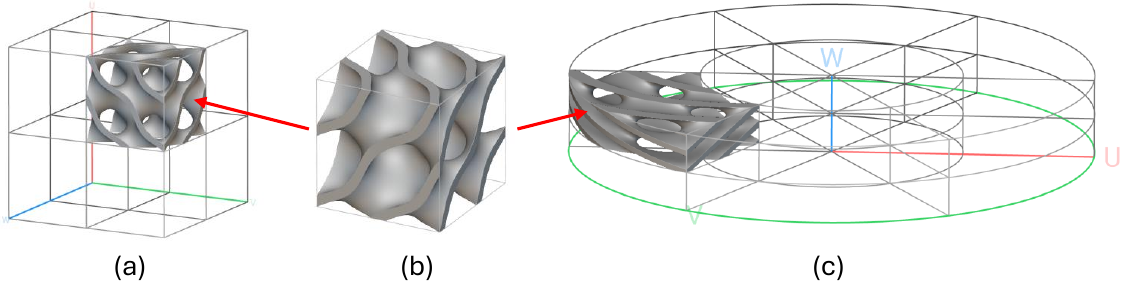}
    \caption{Exemplar cell maps and unit cell: \textbf{a)} Cubic $2\times2\times2$ cell map, cells arranged by Cartesian coordinates, based on cell size XYZ; \textbf{b)} Unit cell diamond-type triply periodic minimal surface (TPMS), implemented into each cell of the chosen cell map, as annotated; \textbf{c)} Cylindrical cell map, cells distorted around polar coordinates, based on cell radius, cell height, and divisions circumferentially, this cell map is more representative of the lattice geometry in this paper.}
    \label{fig:cellmaps}
\end{figure}

To streamline this process, previous in-house lattice selections have been taken forward, building upon the work from \textit{Westsik et al. 2023}\cite{Westsik23}, \textit{Aziz et al. 2025}\cite{Aziz25}, and \textit{Lister et al. 2024}\cite{Lister24} to define the lattice unit cell and cell map. The unit cell is the repeating pattern that forms the lattice structure. A diamond triply periodic minimal surface (TPMS) unit cell was selected as it has been shown to provide the greatest overall stiffness among 32 other unit cell candidates\cite{Westsik23, Lister24, Aziz25}. The cell map is a spatial field used to distribute the unit cell throughout the lattice region. Figure \ref{fig:cellmaps} shows two cell map types and their implementation of a diamond TPMS unit cell. A cubic cell map based on a Cartesian XYZ grid is commonly used due to its simplicity and uniform scaling, as seen in Figure \ref{fig:cellmaps} \textit{(a)}. However, to match the mirror geometry, a cylindrical cell map is used, in which the unit cells are mapped around the central axis using a cylindrical-based coordinate system defined by radial position, height, and circumferential divisions, as shown in Figure \ref{fig:cellmaps} \textit{(c)}.

\begin{figure}
    \centering
    \includegraphics[width=1\linewidth]{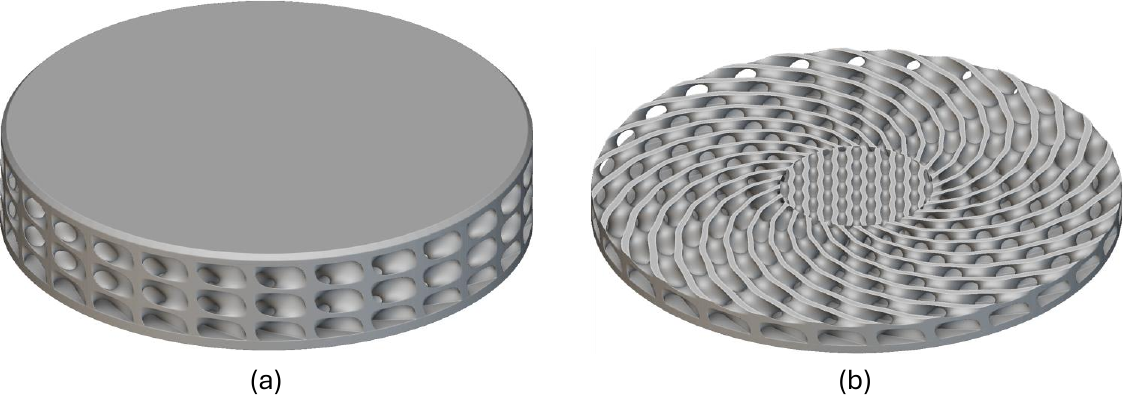}
    \caption{Initial mirror design: \textbf{a)} Full mirror; \textbf{b)} Section view of \textbf{(a)} on-axis to the optical surface, shows the outer lattice mapped to the cylindrical cell map defined in Figure \ref{fig:cellmaps} \textit{(c)} with a wall thickness of \SI{1}{\mm}, the inner lattice follows the cubic cell map defined in Figure \ref{fig:cellmaps} \textit{(a)}, with a wall thickness of \SI{0.8}{\mm}.}
    \label{fig:initialmirror}
\end{figure}

A key limitation of using a cylindrical cell map is the convergence of the lattice towards the centre axis, which leads to local over- and under-support causing stiffness concentrations, resulting in inefficient material distribution when achieving the target mass reduction. To address this, the solution proposed by \textit{Lister et al. 2024}\cite{Lister24} is applied, where the central region of the lattice (the inner lattice) is redefined using a cubic cell map and the remaining lattice (the outer lattice) uses the cylindrical cell map, as shown in Figure \ref{fig:initialmirror}.

\begin{figure}
    \centering
    \includegraphics[width=1\linewidth]{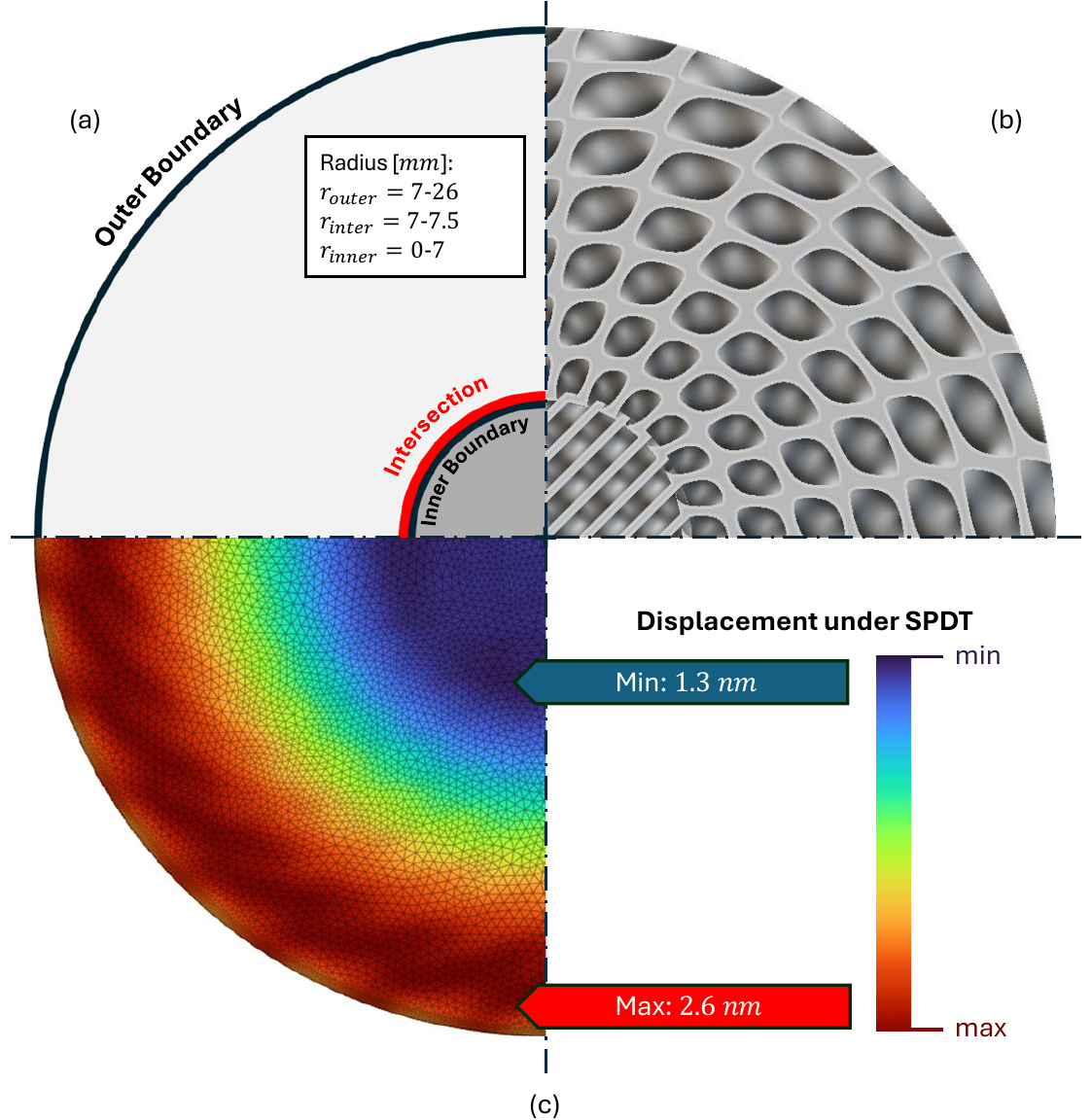}
    \caption{Initial mirror design in quadrants: \textbf{a)} Top-left quadrant, shows boundaries for the inner and outer lattice and their intersection; \textbf{b)} Top-right quadrant, section view parallel to optical surface at \SI{5}{\mm} depth revealing lattice composition, with the outer lattice following cylindrical cell map and the inner lattice following cubic cell map; \textbf{c)} Displacement result from FEA of the optical surface -- base of mirror restrained, optical surface under SPDT pressure \SI{3500}{\Pa}. Centre of the optical surface over-supported with over \SI{1}{\nm} less displacement compared to outer edge of the optical surface.}
    \label{fig:innerouterlatticegeometry}
\end{figure}

The inner lattice extends to a radius of \SI{7.5}{\mm}, with the outer lattice beginning at \SI{7}{\mm}, creating an overlap region (referred to as the intersection) from radii \SIrange{7}{7.5}{\mm}. Beyond \SI{7.5}{\mm}, only the outer lattice is present, as shown in Figure \ref{fig:innerouterlatticegeometry} \textit{(a)}. This arrangement is illustrated in Figure \ref{fig:innerouterlatticegeometry} \textit{(b)}, which shows a sectioned view parallel to the optical surface at \SI{5}{\mm} depth. A displacement-based finite element analysis (FEA) of the optical surface under \SI{3500}{\Pa} pressure load (heritage pressure for SPDT) is seen in Figure \ref{fig:innerouterlatticegeometry} \textit{(c)}. The results show that the inner lattice maintains a relatively uniform displacement response, whereas the outer lattice has a radial change in surface displacement, with over-support near the interface and under-support towards the outer edge. This non-uniform displacement requires material to be redistributed from over-supported regions to under-supported regions to meet design objective \ref{itm:desobj3}.

\begin{figure}
    \centering
    \includegraphics[width=1\linewidth]{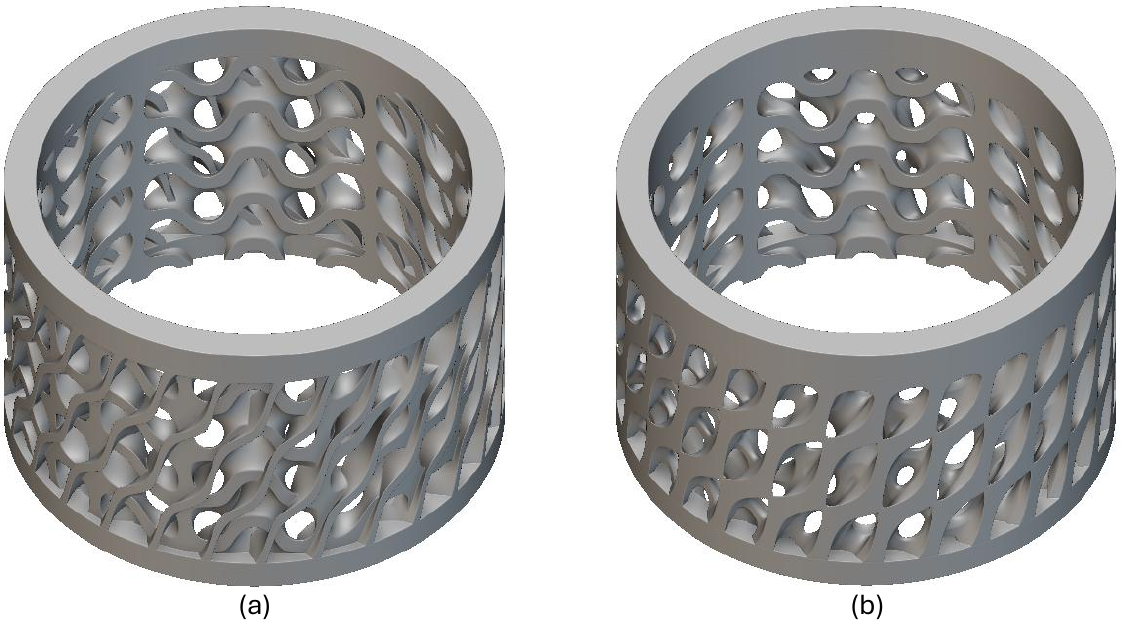}
    \caption{Intersection optimisation: \textbf{a)} intersection before blend; \textbf{b)} intersection after blend.}
    \label{fig:intersection}
\end{figure}

The intersection is optimised by using a blend function within the CAD software, which generates a smooth geometric crossover between the inner and outer lattice. Without blending, the intersection of dissimilar lattice geometries can produce enclosed pockets where overlapping struts meet. The blend function is visualised in Figure \ref{fig:intersection} \textit{(b)} which shows the intersection of the mirror and smooth transition between the inner and outer lattice, while meeting the target mass reduction.

\begin{figure}
    \centering
    \includegraphics[width=1\linewidth]{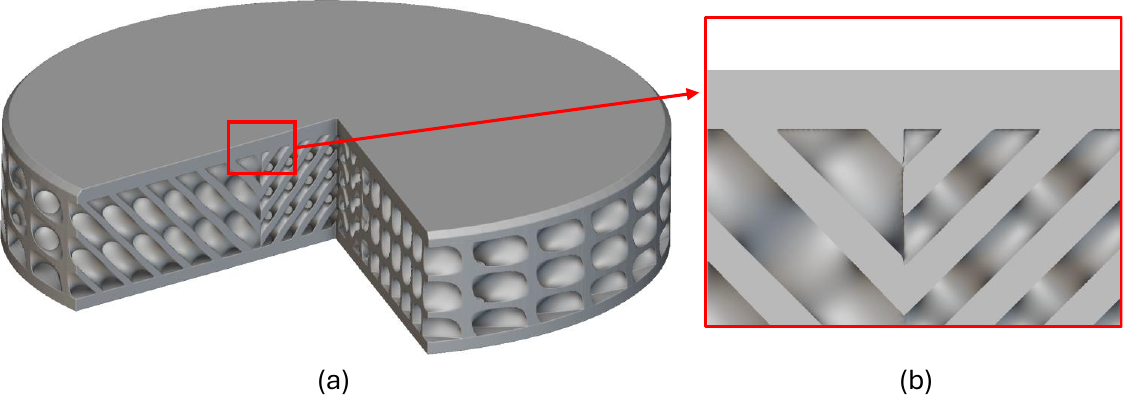}
    \caption{Initial mirror design, slice shows overhang at interface between the lattice and face plate: \textbf{a)} section view of initial mirror design in Figure \ref{fig:initialmirror} \textit{(a)}, highlighting a large horizontal overhang at the interface between the lattice and face plate; \textbf{b)} Expanded view on overhang.}
    \label{fig:InitialOverhang}
\end{figure}

A further opportunity for optimising the mirror design is at the interface between the lattice and face plate. As seen in Figure \ref{fig:InitialOverhang}, the gaps between the lattice walls (approx. \SI{2}{\mm}) form an overhang on the downward facing surface, as highlighted in Figure \ref{fig:InitialOverhang} \textit{(b)}. This overhang must be minimised to meet DfAM requirements (no large overhangs over 45 degrees).

\subsection{Lattice Optimisation}

Following the lattice implementation and evaluation using CAD and FEA in Section \ref{subsec:latticeselection}, two opportunities for further optimisation are identified to meet the design objectives outlined in Section \ref{subsec:desobjs}:

\begin{enumerate}
    \item The lattice material shall be redistributed (maintaining the target mass reduction) to ensure a constant displacement across the optical surface when under SPDT, meeting design objectives \ref{itm:desobj1} and \ref{itm:desobj3}.
    \item Identify and apply solution to minimise overhang at interface between the lattice and face plate, meeting design objective \ref{itm:desobj2}.
\end{enumerate}

To explore a wide range of lattice design variations, a field-driven design (FDD) approach is used\cite{nTopFDD}. FDD allows spatial control of design parameters; given the unit cell, cell map, lattice type, and lattice region parameters are constant, the wall thickness (WT) is the only parameter of the lattice that can be altered in FDD. The WT defines the thickness of the material forming the lattice geometry. For TPMS lattices, since the geometry is initially represented by a zero-thickness mathematical surface, the defined WT value is offset equally on either side of the surface to generate the solid lattice structure. However, within an FDD workflow, spatial control of the WT allows optimisation of material distribution throughout the lattice to meet specific design objectives.

\begin{figure}
    \centering
    \includegraphics[width=1\linewidth]{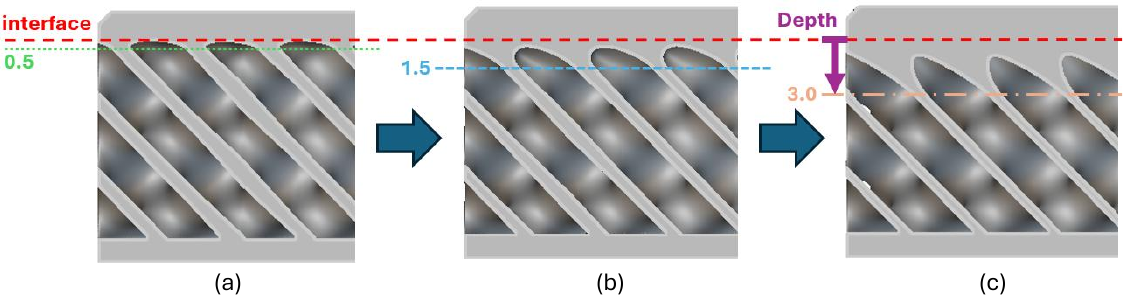}
    \caption{The lattice design approach to minimise overhangs to meet DfAM requirements (Section \ref{sec:specification}): WT = wall thickness, interface is where the lattice and face plate meet, causing an overhang as identified in Figure \ref{fig:InitialOverhang}. To minimise this overhang, the lattice WT is set to gradually increase from a defined depth up to the interface, where at the interface the lattice WT causes adjacent walls to merge locally to minimise unsupported regions. To maintain the target mass of the mirror (50\% mass reduction from solid equivalent), the additional material is taken from the lattice volume, reducing the base WT elsewhere in the structure. This requires iterative adjustment as a reduction in base WT affects the overhang reduction.}
    \label{fig:overhangfixdiagram}
\end{figure}

A proposed depth-based FDD solution for overhang minimisation is seen in Figure \ref{fig:overhangfixdiagram}. Here, the lattice WT is set to gradually increase from a defined depth up to the interface. The rate increase is sufficient that adjacent walls meet at the interface to reduce the overhang size. For example, in Figure \ref{fig:overhangfixdiagram} \textit{(a)}, this begins \SI{0.5}{\mm} below the interface causing a rapid increase. Figure \ref{fig:overhangfixdiagram} \textit{(b)} and \textit{(c)} apply larger depth values of \SI{1.5}{\mm} and \SI{3.0}{\mm} respectively, causing a more gradual increase in lattice WT up to the interface. To maintain the target mass reduction, the material used is taken from the entire lattice by reducing the base WT value. Therefore, this solution requires iterative adjustments as a reduction in the base WT affects the increase throughout the depth-based increase.

\begin{figure}
    \centering
    \includegraphics[width=1\linewidth]{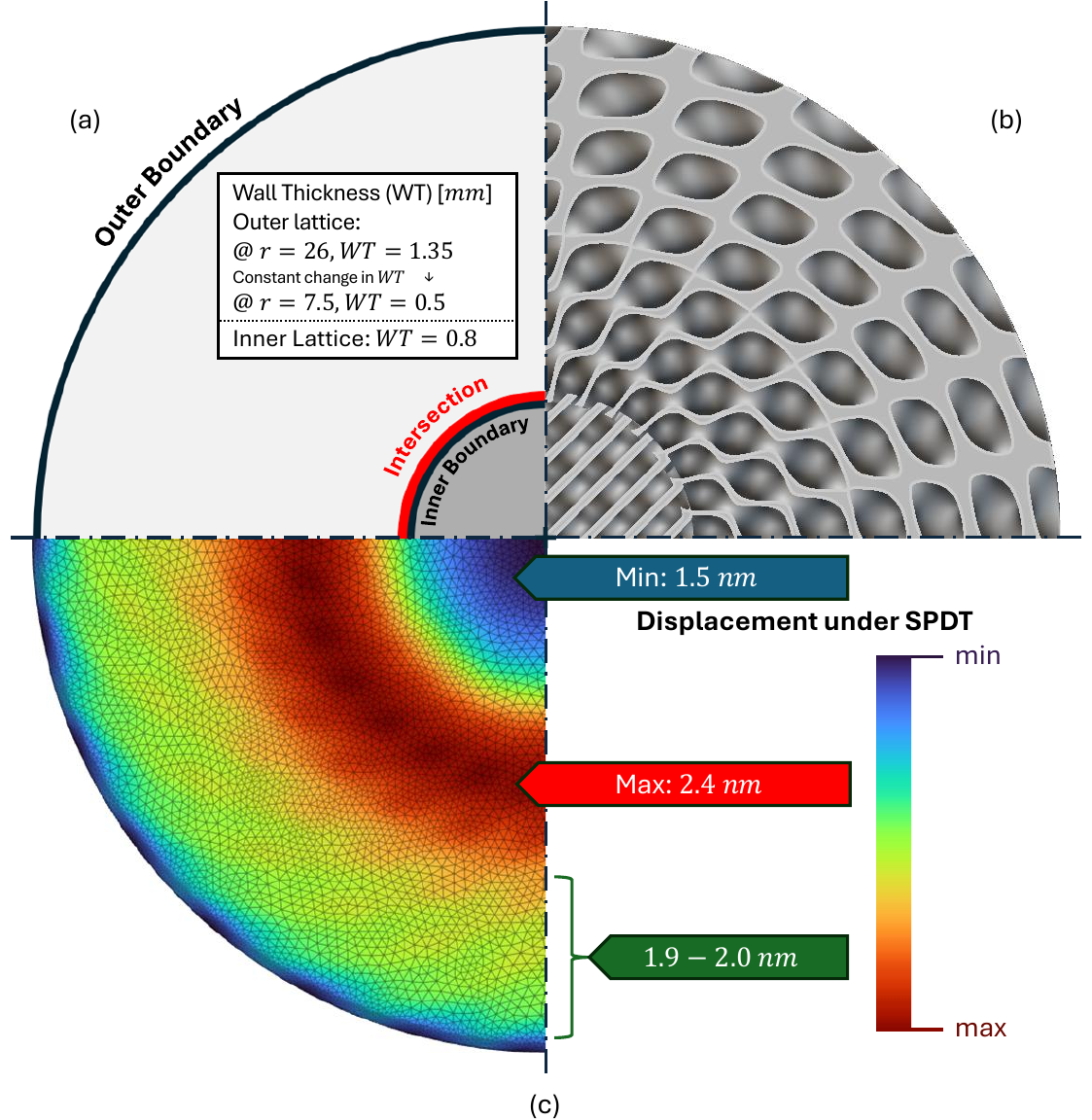}
    \caption{FEA SPDT on radius-based FDD lattice WT mirror, with a linear change in WT from \SIrange{0.50}{1.35}{\mm} from intersection to outer boundary. Results indicate improvement in displacement distribution near outer boundary, however material at intersection requires redistribution to further balance the displacement under SPDT.}
    \label{fig:radiallatticeopt1}
\end{figure}

As there is non-uniform displacement observed in the outer lattice region under FEA SPDT, a radius-based FDD approach is applied. Initially, the WT is defined with a thickness value at the interface and outer boundary respectively, with a linear gradient in WT between these limits. The WT value at the interface is \SI{0.50}{\mm} and \SI{1.35}{\mm} at the outer boundary, a combination that ensures the target mass reduction is maintained while redistributing material from the over-supported interface to the under-supported outer boundary. A repeat FEA test on this mirror design is seen in Figure \ref{fig:radiallatticeopt1}. Here, the intersection is now under-supported with a maximum displacement of \SI{2}{\nm} under SPDT pressure, meaning too much material has been removed from this position of the lattice. However, from a radius of \SIrange{14}{26}{\mm}, the displacement is maintained at around \SIrange{1.9}{2.0}{\nm} which meets design objective \ref{itm:desobj3}. As the centre remains over-supported, it is clear that iterations on the radius-based FDD approach is required to achieve a mirror with a uniform displacement across the optical surface when under SPDT pressure.

\begin{figure}
    \centering
    \includegraphics[width=0.8\linewidth]{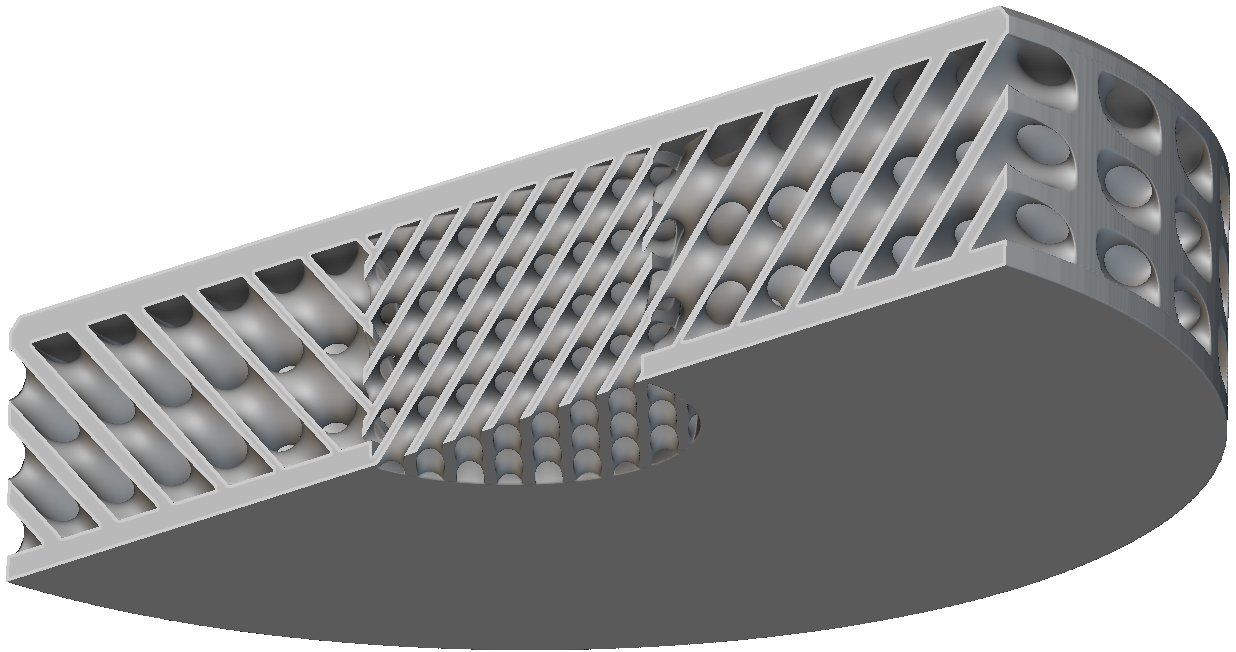}
    \caption{Section view of initial mirror design: added extension of the inner lattice through the base plate to relocate material to the outer lattice for uniform displacement across the optical surface.}
    \label{fig:innerlatticeextension}
\end{figure}

To satisfy both the overhang minimisation and surface displacement objective, the depth-based and radius-based FDD solutions were combined into a single FDD approach. Although each solution is simple to optimise independently while maintaining the target mass reduction, doing so would fully allocate the available material to a single objective so there would be insufficient material to implement the second FDD solution without exceeding the WT constraint. By combining both FDD solutions, the available material could be distributed simultaneously and allow both objectives to be evaluated under the same target mass reduction. Here, the single FDD solution varies WT as a function of both spatial depth and radius, enabling an iterative approach to find an ideal mirror design for laser remelting. Additionally, given the inner lattice displayed a constant displacement across the optical surface under SPDT and was over-supported, the lattice was extended through the base of the mirror to relocate the material from the inner lattice to the outer lattice. This extension is shown in Figure \ref{fig:innerlatticeextension}.

\begin{figure}
    \centering
    \includegraphics[width=1\linewidth]{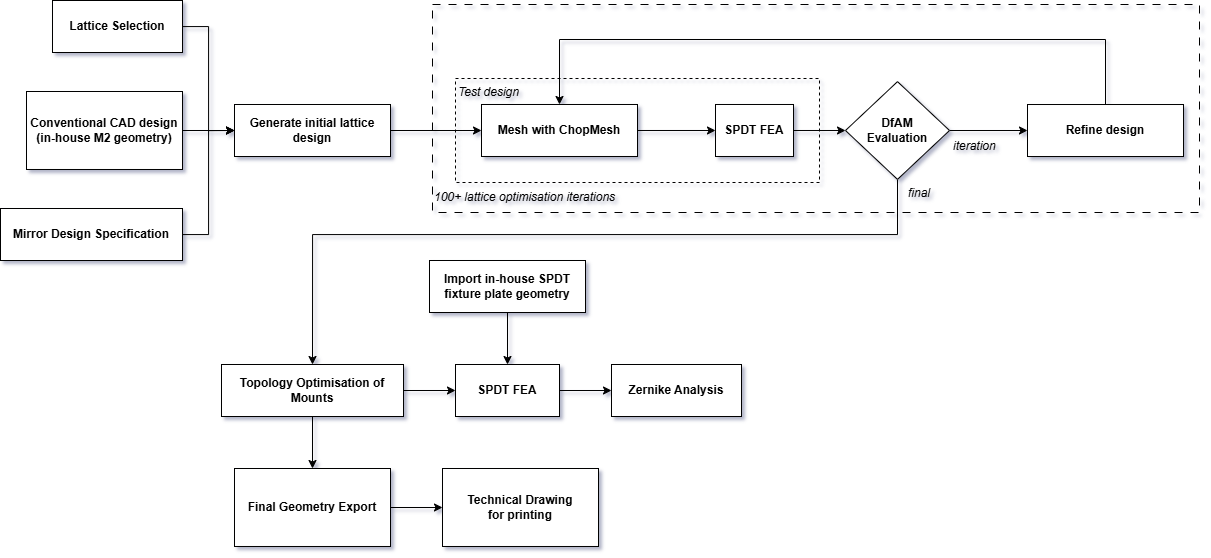}
    \caption{Iterative process followed to generate a mirror design.}
    \label{fig:DesignChain}
\end{figure}

The use of this FDD introduces a large and continuous design space, where small changes can produce unique lattice configurations. Consequently, the design space becomes too large for manual investigation alone. To accelerate the iteration process and ensure a broad range of lattice variations were explored, two initial radius-based WT ratios were selected for the outer lattice, with WTs of $4:3$ or $3:2$ between the outer boundary (max) and intersection (min). Furthermore, WT thicken depths between \SI{1.0}{\mm} and \SI{2.5}{\mm} were investigated at \SI{0.5}{\mm} increments. Figure \ref{fig:DesignChain} depicts the overall lattice optimisation method, in which 68 lattice configurations that follow the predefined lattice WT ratios, thicken depth increments, and target mass reduction are tested. All iterations underwent FEA SPDT to test the displacement distribution at the optical surface for design objective \ref{itm:desobj3}. A downselection of 8 initial FDD lattice WT configurations were taken forward, with their parameters listed in Table \ref{tab:FDDLatticeDesigns}. 

{\renewcommand{\arraystretch}{1.25}
\begin{table}[h]
\centering
\caption{Initial FDD lattice WT configurations based on max and min WT, and thicken depth while keeping mass reduction above 50\% -- therefore, an increase in thicken depth requires the reduction in max and min WT (consistent to the WT ratio).}
\newcolumntype{C}[1]{>{\centering\arraybackslash}m{#1}}
\begin{tabular}{|C{1.2cm}||C{1.8cm}|C{2.3cm}|C{2.3cm}|C{2.3cm}|C{1.5cm}|}
\hline
\textbf{Lattice} &
\textbf{WT Ratio} &
\textbf{Max WT (mm)} &
\textbf{Min WT (mm)} &
\textbf{Thicken Depth (mm)} &
\textbf{Mass reduction (\%)} \\
\hline
1 & \multirow{4}{*}{4:3} & 1.05 & 0.79 & 1.0 & 50.12 \\
2 &  & 0.95 & 0.71 & 1.5 & 50.00 \\
3 &  & 0.88 & 0.66 & 2.0 & 49.97 \\
4 &  & 0.80 & 0.60 & 2.5 & 50.08 \\
\hline
5 & \multirow{4}{*}{3:2} & 1.10 & 0.70 & 1.0 & 50.02 \\
6 &  & 0.98 & 0.65 & 1.5 & 50.04 \\
7 &  & 0.93 & 0.62 & 2.0 & 50.04 \\
8 &  & 0.84 & 0.56 & 2.5 & 50.13 \\
\hline
\end{tabular}
\label{tab:FDDLatticeDesigns}
\end{table}}

\begin{figure}
    \centering
    \includegraphics[width=0.5\linewidth]{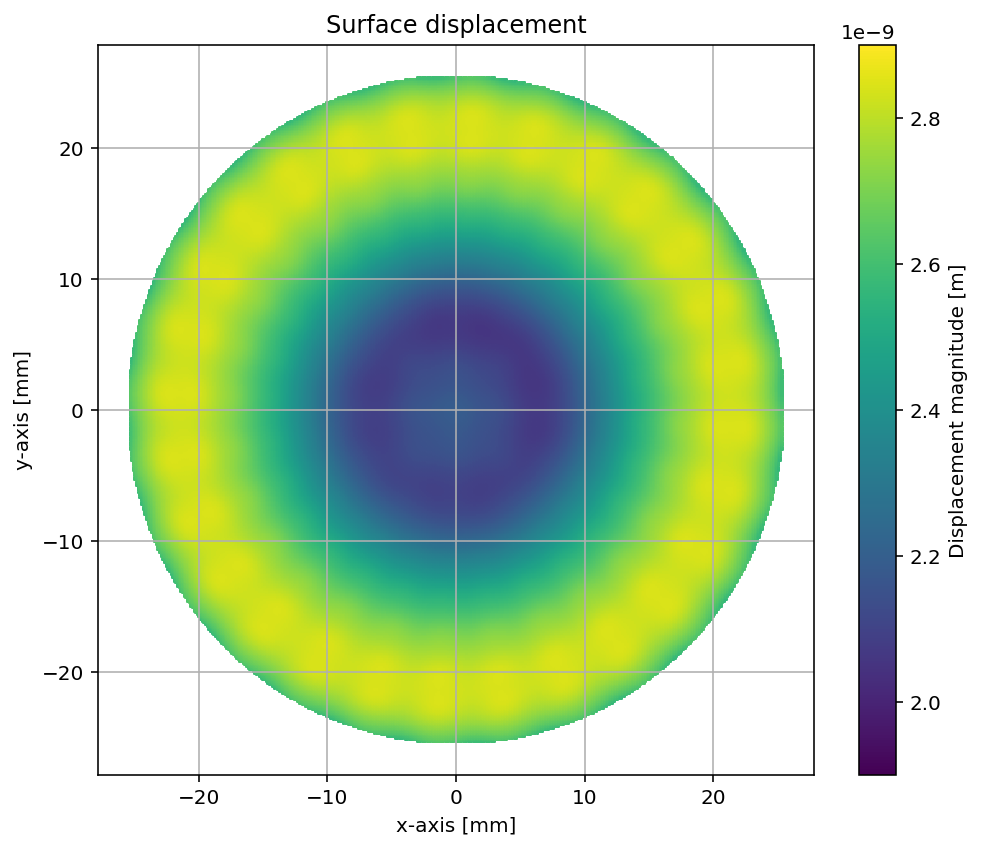}
    \caption{Down selected FDD lattice WT configuration under FEA SPDT, regions remain over-supported and further optimisation is required.}
    \label{fig:latticedownselect}
\end{figure}

\begin{figure}
    \centering
    \includegraphics[width=0.7\linewidth]{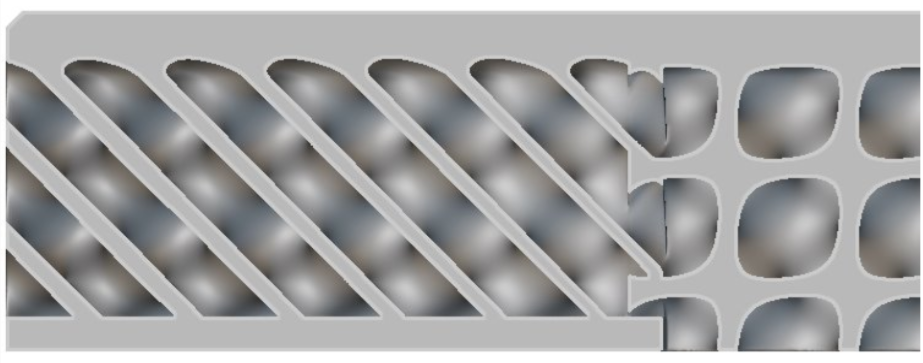}
    \caption{Final lattice design, overhang minimised at intersection and material redistributed sufficiently to ensure uniform displacement at the optical surface under SPDT.}
    \label{fig:latticefinal1}
\end{figure}

\begin{figure}
    \centering
    \includegraphics[width=0.5\linewidth]{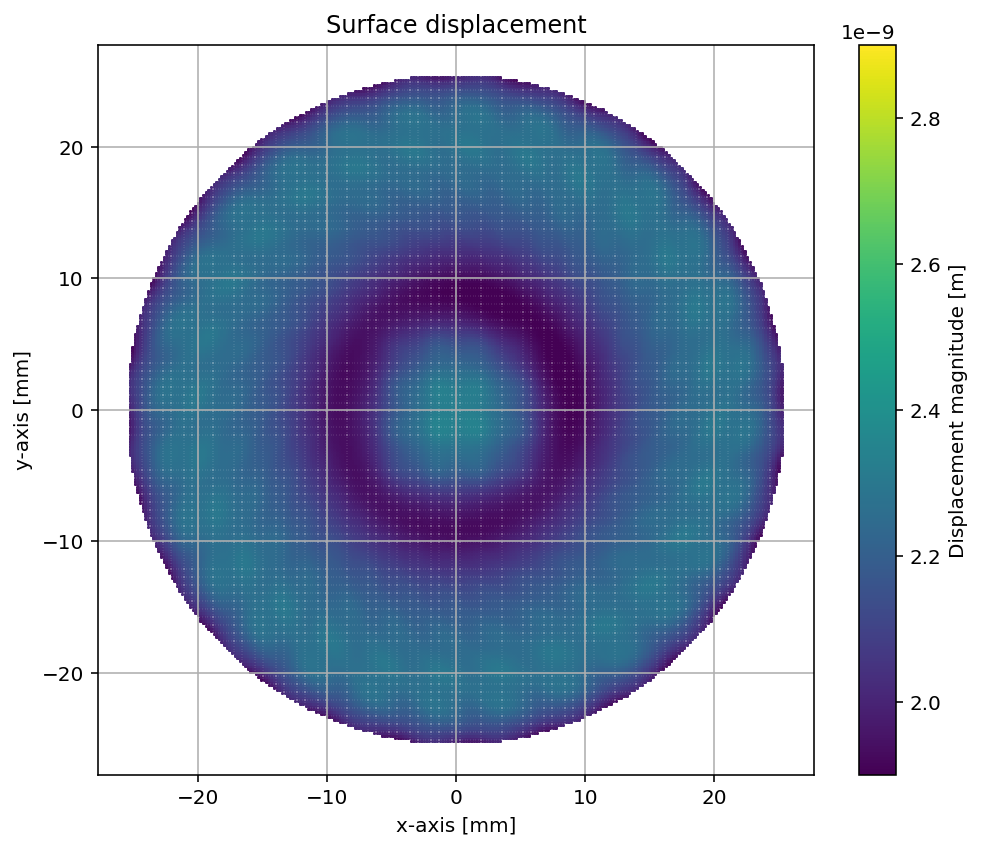}
    \caption{Final lattice design, displacement under FEA: Colour bar matches Figure \ref{fig:latticedownselect} for comparison. Uniform displacement at surface is now based on lattice geometry beneath dominating the displacement variation.}
    \label{fig:latticefinal2}
\end{figure}

All configurations sufficiently minimised the overhang, as adjacent walls of the TPMS lattice blended at the interface, reducing the overhang angle and complying with DfAM rules (maximum \SI{45}{\degree}). Therefore, to maximise the available material to balance and reduce displacement under SPDT, a thicken depth of \SI{1.0}{\mm} is taken forward for further designs, leaving lattice 1 or 5. Based on performance in FEA, the displacement distribution of lattice 5 was more uniform than lattice 1, so is taken forward. As seen in Figure \ref{fig:latticedownselect}, lattice 5 continues to present over-support and under-support at the optical surface when under simulated SPDT, where the design only implements a linear change between max and min WT. Therefore, manual optimisation of the lattice is conducted, where instead of a linear change in WT, specific radii values are selected to relocate material. Further iterations were required to optimise the displacement distribution, with an acceptable final design seen in Figure \ref{fig:latticefinal1} and its associated distribution of FEA SPDT seen in Figure \ref{fig:latticefinal2}. Here, the displacement is uniform across the optical surface, with the lattice geometry beneath the mirror dominating the displacement variation.

\subsection{Mounting features}

Based on in-house mount design experience, a mounting solution proposed by \textit{Lister et al. 2024}\cite{Lister24} is taken forward as the mirror dimensions are comparable. This solution uses topology optimisation (TO) to reduce mass on the mounting features of a mirror. TO is a computational design approach used, in this case, to minimise mass usage to 50\% compared to a solid equivalent while supporting the \SI{3500}{\Pa} SPDT pressure. The boundary conditions were set to ensure the mounting holes produced are at least \diameter \SI{10}{\mm} in the defined locations from Table \ref{tab:DesSpec}. As the geometry is only constrained by the mounting hole locations, the intermediate volume toward the mirror is defined as the design domain. Starting from a solid design space, the optimisation process iteratively removes material that contributes least to load-bearing under SPDT while maintaining load transfer efficiency, resulting in an efficient structural pathway between constraints. The output of TO is a mount with reduced mass while maintaining required stiffness and strength for SPDT.

\begin{figure}
    \centering
    \includegraphics[width=1\linewidth]{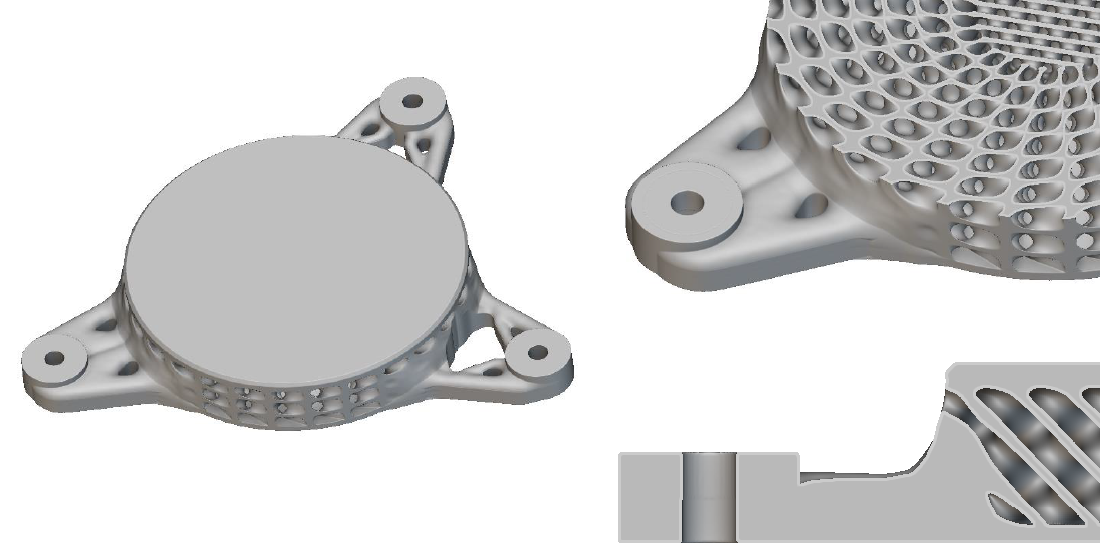}
    \caption{Topology optimised mount implemented to mirror design.}
    \label{fig:mount1}
\end{figure}

As a sandwich mirror design is used, the side walls of the mirror that would typically be used to support mounting features are absent. Therefore, the mount is blended into the lower half of the lattice to ensure the mounts are secure to the mirror and do not affect the optical surface through induced stresses. Figure \ref{fig:mount1} shows the TO mount implemented to the mirror at the designated locations based on the design specification.

\begin{figure}
    \centering
    \includegraphics[width=1\linewidth]{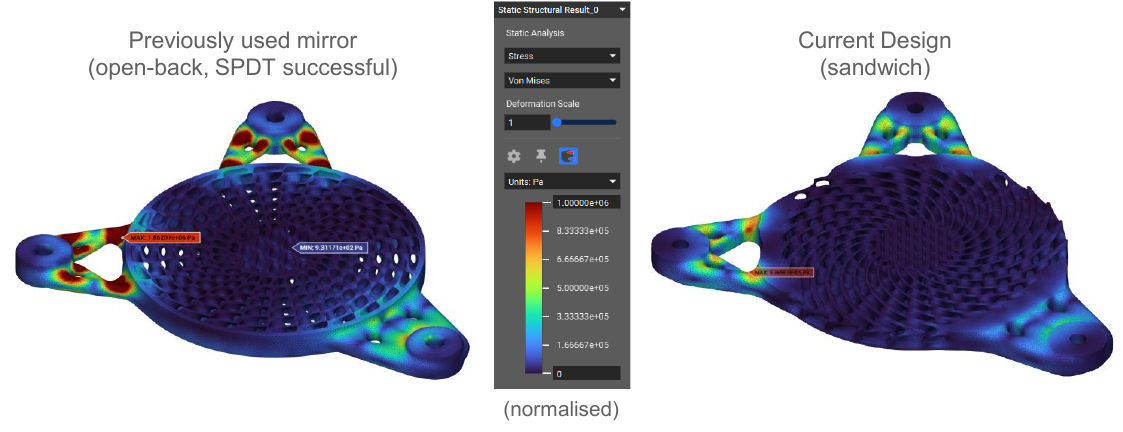}
    \caption{Verification of mounts through FEA SPDT to test for Von Mises stress.}
    \label{fig:mountcheck}
\end{figure}

The mounts are verified by FEA SPDT to measure the Von Mises stress induced within the mounts. As seen in Figure \ref{fig:mountcheck}, the current mount design exhibits lower stress levels than the mounting solution proposed by \textit{Lister et al. 2024}\cite{Lister24}. It should be noted that the primary design focus was DfAM rather than mount optimisation, so the mounting features are not directly equivalent, as the current design is extruded and embedded within the lattice structure to facilitate powder removal and enable the laser remelting process. Despite this difference, the reduced stress observed in the current design indicates that the mounting features are structurally acceptable.

\section{Mirror fabrication and Metrology}
\label{sec:manufacture}

\begin{figure}
    \centering
    \includegraphics[width=1\linewidth]{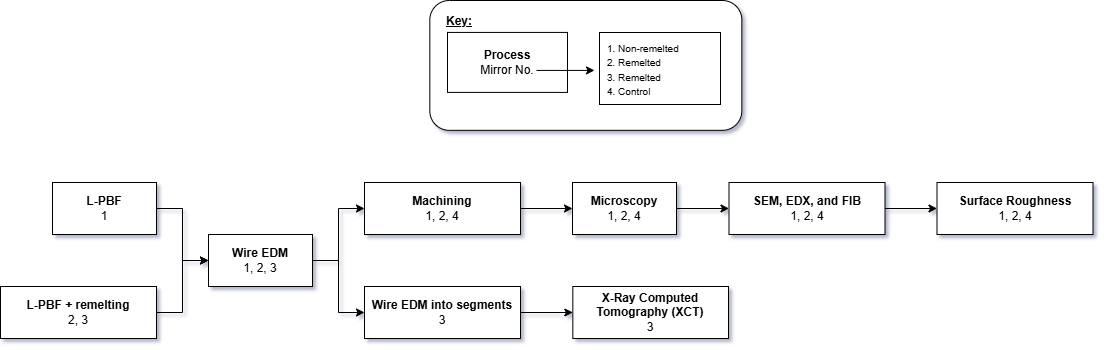}
    \caption{Process chain followed for manufacturing and metrology for mirrors.}
    \label{fig:machiningandmetrology}
\end{figure}

\begin{figure}
    \centering
    \includegraphics[width=1.0\linewidth]{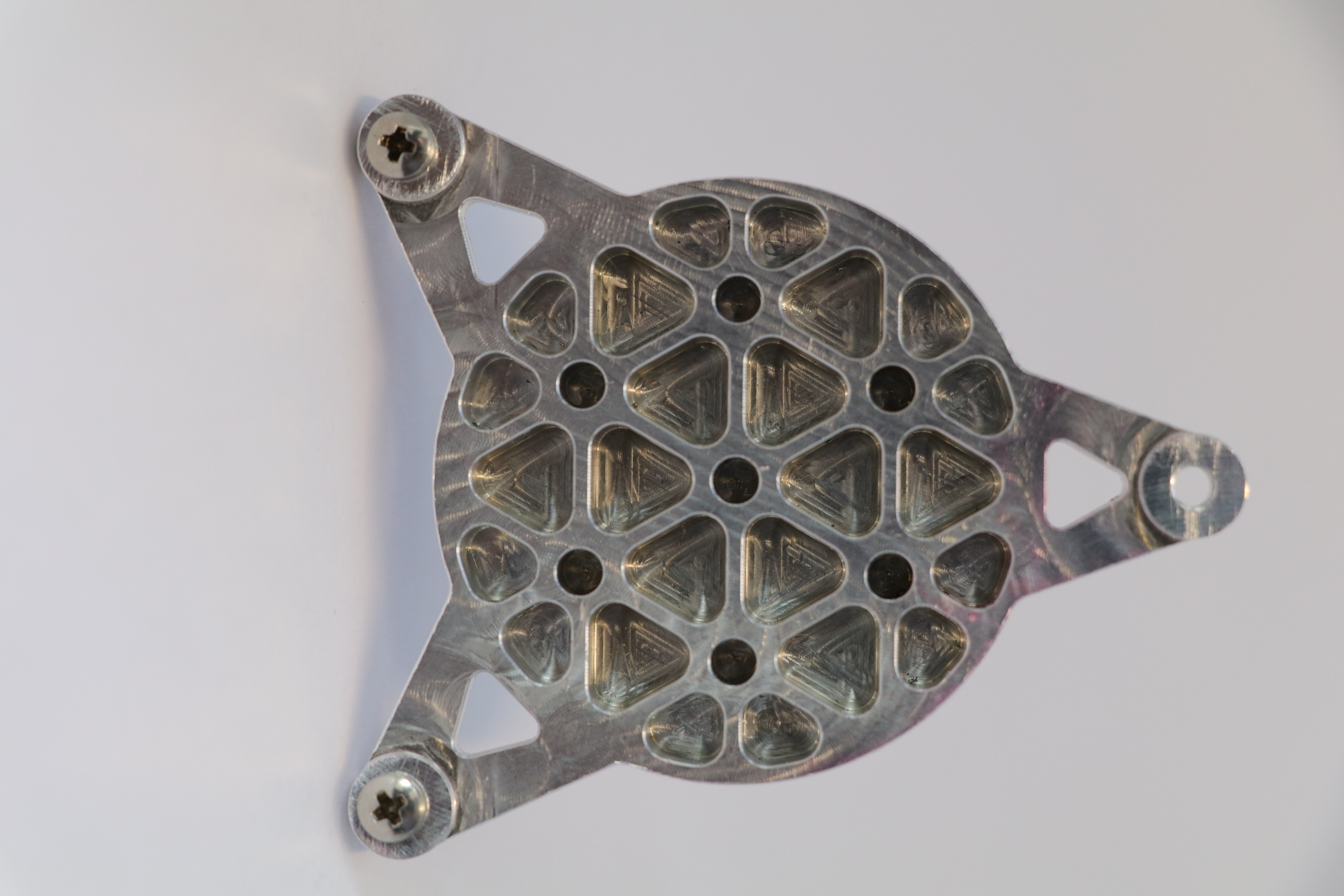}
    \caption{Control mirror, machined from an RSA 6061 billet. Open-back design with an iso-grid pattern milled from the base for \SI{55}{\%} mass reduction. Image credit Jason Cowan.}
    \label{fig:controlmirror}
\end{figure}

With the final mirror design, the process chain Figure \ref{fig:machiningandmetrology} is followed for the fabrication and metrology of the first set of mirrors. Two of these mirrors are manufactured with laser remelting while a third is made without for comparison. Furthermore, a mirror conventionally manufactured via machining from an RSA 6061 billet to the same dimensions is made to act as a control. This mirror has an open-back design with \SI{30}{\g} mass reduction from a \SI{55}{\g} solid equivalent (excluding mounts) by milling an iso-grid pattern from the base, as seen in Figure \ref{fig:controlmirror}. The addition of mounts introduces \SI{5}{\g} to the total mass.

\subsection{Additive Manufacturing}

Based on the parameter screening conducted at KU Leuven with R2\_9 showing ideal results in terms of porosity mitigation, the manufacturing parameters for the AM mirrors are shown in Table \ref{tab:printparameters}.

{\renewcommand{\arraystretch}{1.25}
\begin{table}[h]
\centering
\caption{Manufacturing parameters for the AM mirrors.}
\newcolumntype{C}[1]{>{\centering\arraybackslash}m{#1}}
\begin{tabular}{|C{2.7cm}||C{1.5cm}|C{2.1cm}|C{1.5cm}|C{2.1cm}|C{1.5cm}|C{2.4cm}|}
\hline
\textbf{\diagbox[width=3.15cm]{\textbf{Process}}{\textbf{Param.}}} & \textbf{Power {[}W{]}} & \textbf{Scan speed {[}mm/s{]}} & \textbf{Hatch {[}$\pmb{\mu}$m{]}} & \textbf{Height step {[}$\pmb{\mu}$m{]}} & \textbf{Passes} & \textbf{Scan rotation {[}deg{]}} \\
\hline
LPBF & 350 & 1000 & 100 & 30 & -- & -- \\
\hline
Laser Remelting & 350 & 700 & 70 & 30 & 2 & 67 \\
\hline
\end{tabular}
\label{tab:printparameters}
\end{table}}

\begin{figure}
    \centering
    \includegraphics[width=1\linewidth]{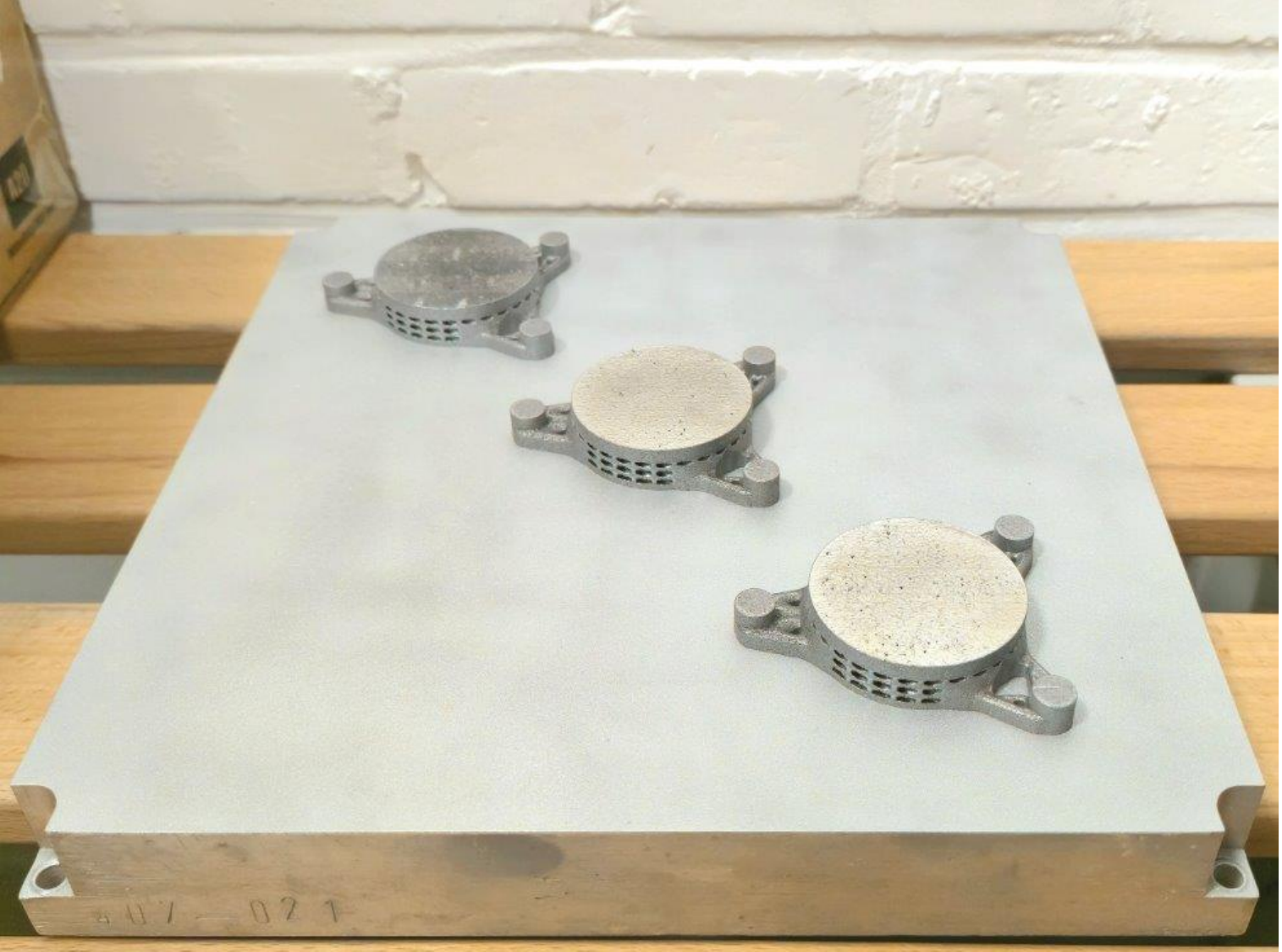}
    \caption{AM mirrors attached to the build plate. Left mirror is the non-remelted, while the centre and right mirrors are remelted; a cosmetic surface finish was applied to distinguish the mirrors. Image credit Sameer Dayanand Meshram (KU Leuven, Belgium).}
    \label{fig:printedonplate}
\end{figure}

Laser remelting was applied for \SI{1.5}{\mm} within the bulk material, \SI{2.25}{\mm} below the as-built top surface, to allow enough buffer zone for the subsequent machining and SPDT post-processes on the mirror to realise the optical surface, while ensuring consistency in the lattice region (between a remelted and non-remelted mirror). An additional \SI{2.25}{\mm} material was added on top of the remelted region and mount faces, \SI{1}{\mm} to the base, and the mounting holes filled to facilitate machining before SPDT, where all additional material is machined off to ensure the mirror geometry meets the design. The mirrors are seen in Figure \ref{fig:printedonplate}, attached to the build plate. These mirrors were removed from the build plate via wire EDM and cleaned to remove the residual powder within the lattice.

\subsection{Post-Processing}

\begin{figure}
    \centering
    \includegraphics[width=1\linewidth]{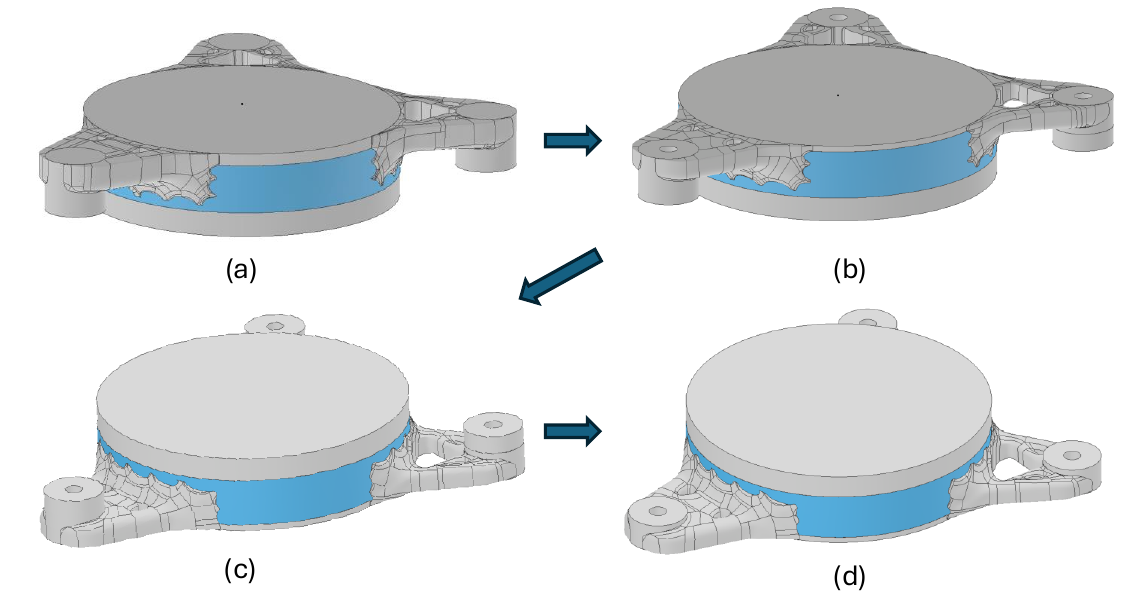}
    \caption{Machining steps to prepare AM mirror for SPDT: \textit{a)} mirror as built; \textit{b)} base machined off to generate datum, followed by mount through holes machined out; \textit{c)} mirror flipped; \textit{d)} final machined mirror, mounting pads and additional material machined down.}
    \label{fig:machiningsteps}
\end{figure}

To evaluate the structural and surface effects of the laser remelting process, these three substrates were allocated to parallel testing tracks:
\begin{enumerate}
    \item One remelted mirror was designated for X-ray computed tomography (XCT) at KU Leuven to inspect the internal pore distribution and subsurface porosity.
    \item The remaining three mirrors (remelted, non-remelted, control) were sent to UK ATC for machining and SPDT to generate a functional optical surface on each. The machining route seen in Figure \ref{fig:machiningsteps} was followed to prepare the mirror for SPDT.
\end{enumerate}

\subsubsection{Single Point Diamond Turning}

\begin{figure}
    \centering
    \includegraphics[width=1\linewidth]{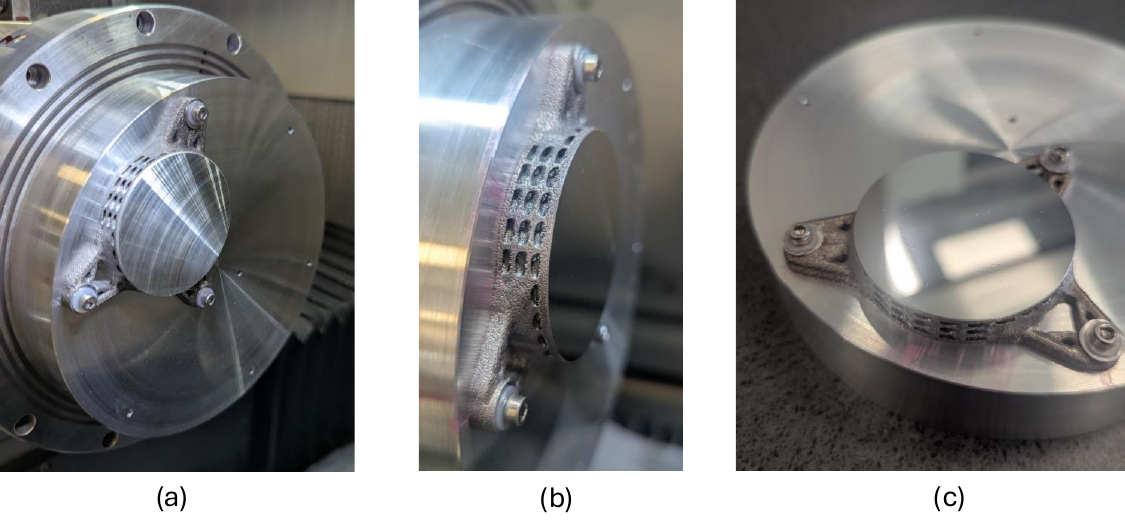}
    \caption{SPDT of remelted mirror: \textit{a)} preparation for SPDT, mirror mounted in CNC lathe off-axis to diamond tool (aligned to centre of SPDT fixture); \textit{b)} Mirror diamond turned; \textit{(c)} Final mirror removed from CNC lathe.}
    \label{fig:SPDTprocess}
\end{figure}

\begin{figure}
    \centering
    \includegraphics[width=1\linewidth]{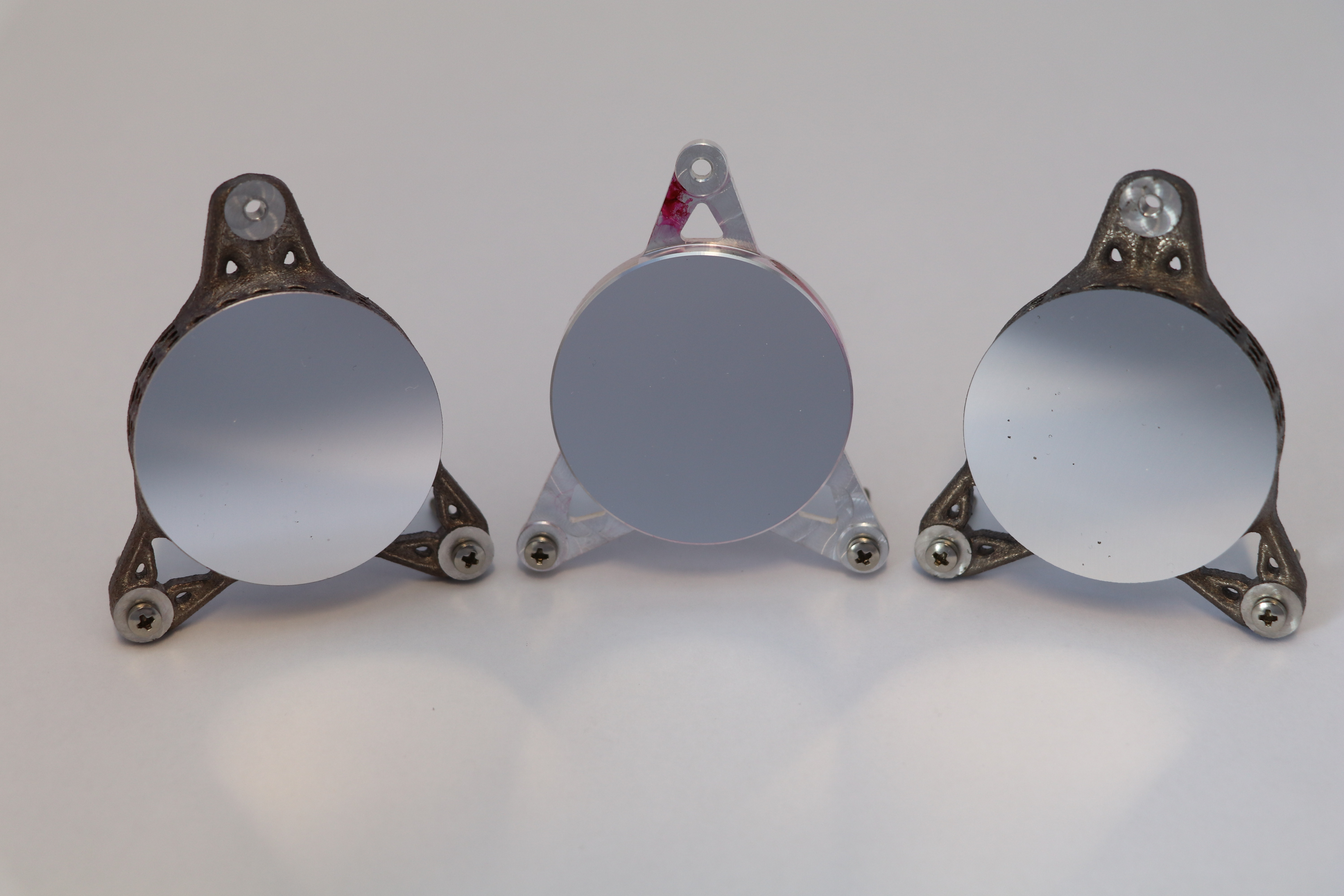}
    \caption{Final mirrors: \textit{left} - non-remelted; \textit{middle} - control; \textit{right} - remelted}
    \label{fig:finalmirrors}
\end{figure}

For the second track, after machining, SPDT was performed at RAL CLF to generate the optical surface. Material removal was conducted through progressively decreasing depths of cut, reducing cutting forces during the final finishing passes and minimising the risk of surface damage. The same SPDT process parameters and mirror arrangement was used for all mirrors to ensure that any observed differences in optical performance could be attributed to the laser remelting process rather than variations in optical fabrication. Figure \ref{fig:SPDTprocess} shows the remelted mirror at three stages in the SPDT process, and the final mirrors with an optical surface are seen in Figure \ref{fig:finalmirrors}.

\subsection{Metrology}

A similar analysis route to the proof-of-concept cubes is conducted, as defined in process chain Figure \ref{fig:machiningandmetrology}. Where optical microscopy, SEM, EDX, FIB, and surface roughness measurements are conducted to characterise the optical surfaces and allow comparison between the remelted, non-remelted, and control mirror. Additionally, KU Leuven performed an Archimedes' principle test to measure the mass of the AM mirrors and estimate the porosity \% in each. 

\begin{figure}
    \centering
    \includegraphics[width=1\linewidth]{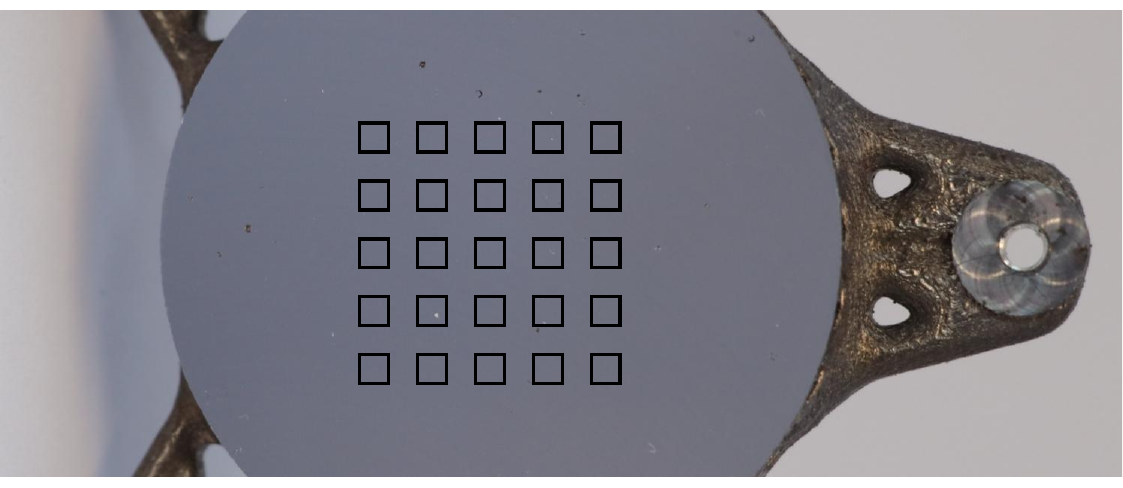}
    \caption{5$\times$5 square arrangement for sampling surface roughness.}
    \label{fig:brukerlayout}
\end{figure}

The surface roughness of each mirror was measured with the Bruker Contour-X 200 with a similar setup as the proof-of-concept cubes - more regions are captured by measuring 25 segments of each optical surface in a 5$\times$5 square arrangement, as shown in Figure \ref{fig:brukerlayout}. Each segment captured provides data on the Mean (Sa), RMS (Sq), and PV (Sz) surface roughness of the region.

\begin{figure}
    \centering
    \includegraphics[width=1\linewidth]{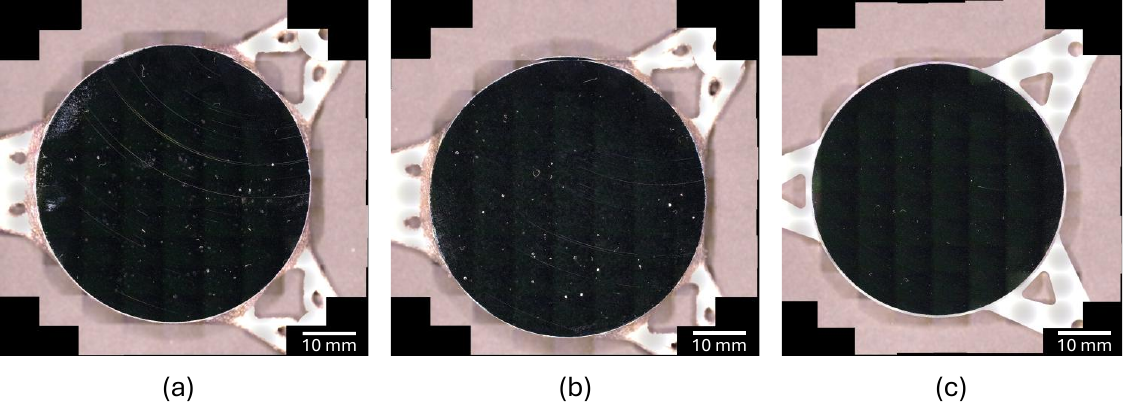}
    \caption{Dark-field images of mirrors: \textit{a)} Non-remelted; \textit{b)} Remelted; \textit{c)} Control.}
    \label{fig:DFoptical}
\end{figure}

One BF and DF image was taken for each optical surface using the Evident DSX2000, requiring a reduced magnification of 3$\times$ to capture the entire surface and slight adjustment to the exposure time with \SI{373}{\us} for the BF images and \SI{285}{\us} for the DF images. The DF images are seen in Figure \ref{fig:DFoptical}, where all surface artefacts are amplified and visible.

\section{Metrology Results and Discussion}
\label{sec:results}

Results from the XCT analysis on one of the remelted mirrors highlight the presence of two populations of defects within the sampled regions: small ($<<$\SI{100}{\um}) highly spherical pores make up the majority of the detected defects, while a population of clear lack-of-fusion pores with circumscribed diameters up to \SI{760}{\um} and a calculated sphericity value $<<$0.5 are identified within the remelted section. While the calculated XCT total porosity content is only 0.11\%, overall, the presence of these large defects can be extremely deleterious for the mirror's optical performance. Furthermore, the Archimedes' Principle test predicted at least 1\% porosity throughout all AM mirrors, as seen in Table \ref{tab:densityresults}. This value is higher than the total porosity content calculated by XCT; this discrepancy can be related to the XCT resolution limit (voxel size of \SI{7}{\um}) and the high variability of Archimedes measurements, which can have considerable systematic errors if not correctly calibrated for the exact densities of the material and medium.

{\renewcommand{\arraystretch}{1.25}
\begin{table}[h]
\centering
\caption{Archimedes' principle test on AM mirrors.}
\newcolumntype{C}[1]{>{\centering\arraybackslash}m{#1}}
\begin{tabular}{|C{2.2cm}||C{2.3cm}|C{2.3cm}|C{2.3cm}|C{2.3cm}|C{2cm}|}
\hline
\textbf{Mirror} & \textbf{Weight in air {[}g{]}} & \textbf{Weight in ethanol {[}g{]}} & \textbf{Density {[}g/cm$^3${]}} & \textbf{Relative density {[}\%{]}} & \textbf{Porosity {[}\%{]}} \\
\hline
Remelted 1 & 66.3835 & 46.583 & 2.6418 & 98.94 & 1.06 \\
\hline
Remelted 2 & 67.6565 & 47.478 & 2.6420 & 98.95 & 1.05 \\
\hline
Non-remelted & 66.9514 & 46.935 & 2.6356 & 98.71 & 1.29 \\
\hline
\end{tabular}
\label{tab:densityresults}
\end{table}}

\begin{figure}
    \centering
    \includegraphics[width=1\linewidth]{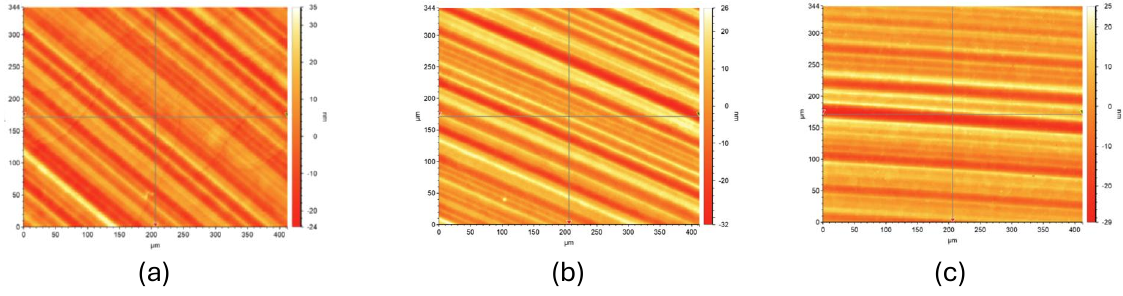}
    \caption{A sampled region of each optical surface, showing surface roughness topography: \textit{a)} Non-remelted; \textit{b)} Remelted; \textit{c)} Control.}
    \label{fig:surfacetopology}
\end{figure}

Table \ref{tab:SurfaceRoughnessMirror} shows the collation of surface roughness data for each mirror. The mean surface roughness RMS (Sq) for all mirrors were measured at higher values than expected, at \SI{11.8}{\nm}, \SI{11.8}{\nm}, and \SI{9.3}{\nm} for the non-remelted, remelted, and control mirror respectively. Given the standard deviation of the surface roughness RMS (Sq) of each mirror, the surface roughness measurements did not demonstrate a clear distinction between the mirrors. Across all sampled regions the measured topography was dominated by the cutting marks of the diamond tool from the SPDT process, which is the most influential feature regardless of the mirror type, as seen in Figure \ref{fig:surfacetopology}. The tooling pattern was consistent across all measured segments for each mirror, indicating that the high surface roughness RMS is caused by the SPDT process rather than subsurface features introduced during AM or conventional manufacturing.

{\renewcommand{\arraystretch}{1.25}
\begin{table}[h]
\centering
\caption{Surface roughness measurements for Non-Remelted, Remelted, and Control mirrors with Mean (Sa), RMS (Sq), and Peak-to-Valley (Sz) metrics across all positions.}
\newcolumntype{C}[1]{>{\centering\arraybackslash}m{#1}}
\begin{tabular}{|m{2.8cm}||C{1cm}|C{1cm}|C{1cm}||C{1cm}|C{1cm}|C{1cm}||C{1cm}|C{1cm}|C{1cm}|}
\hline
\multirow{2}{*}{\diagbox[width=3.25cm,height=1.08cm]{\textbf{Mirror}}{\textbf{Position}}} & \multicolumn{3}{c||}{\textbf{Non-Remelted {[}nm{]}}} & \multicolumn{3}{c||}{\textbf{Remelted {[}nm{]}}} & \multicolumn{3}{c|}{\textbf{Control {[}nm{]}}} \\
\cline{2-10}
& \textbf{Sa} & \textbf{Sq} & \textbf{Sz} & \textbf{Sa} & \textbf{Sq} & \textbf{Sz} & \textbf{Sa} & \textbf{Sq} & \textbf{Sz} \\
\hline
Position 1  & 8.3 & 11.2 & 199 & 6.8 & 9.3 & 71.8 & 9.5 & 12.4 & 75.0 \\
Position 2  & 8.8 & 11.4 & 88.3 & 6.9 & 8.6 & 71.2 & 5.5 & 7.2 & 178 \\
Position 3  & 10.6 & 12.9 & 68.3 & 9.8 & 11.9 & 83.1 & 6.2 & 7.9 & 144 \\
Position 4  & 11.4 & 13.4 & 159 & 11.5 & 14.5 & 107 & 8.4 & 10.0 & 156 \\
Position 5  & 8.6 & 10.6 & 120 & 10.6 & 13.2 & 261 & 7.0 & 9.2 & 276 \\
Position 6  & 10.1 & 12.6 & 102 & 8.2 & 10.8 & 73.6 & 6.5 & 7.9 & 115 \\
Position 7  & 8.5 & 10.7 & 89.6 & 8.5 & 10.4 & 56.9 & 7.2 & 8.8 & 131 \\
Position 8  & 9.6 & 12.6 & 128 & 8.7 & 10.8 & 132 & 6.8 & 8.6 & 80.8 \\
Position 9  & 9.6 & 11.9 & 135 & 9.2 & 11.1 & 63.9 & 7.9 & 17.4 & 859 \\
Position 10 & 9.6 & 11.8 & 79.4 & 8.7 & 10.6 & 58.0 & 5.9 & 7.5 & 108 \\
Position 11 & 7.1 & 9.9 & 109 & 10.0 & 14.5 & 1154 & 6.9 & 8.6 & 267 \\
Position 12 & 8.7 & 11.2 & 151 & 11.1 & 15.0 & 272 & 7.1 & 9.3 & 209 \\
Position 13 & 8.9 & 13.2 & 59 & 8.8 & 10.8 & 58.8 & 6.0 & 7.3 & 160 \\
Position 14 & 7.5 & 9.5 & 58.1 & 8.9 & 11.1 & 79.9 & 6.6 & 8.6 & 67.1 \\
Position 15 & 9.3 & 11.5 & 98.5 & 9.3 & 11.7 & 85.7 & 7.1 & 8.8 & 229 \\
Position 16 & 9.7 & 12.9 & 286 & 9.1 & 11.1 & 135 & 8.0 & 10.9 & 91.0 \\
Position 17 & 7.6 & 9.0 & 58.6 & 11.2 & 14.0 & 82.0 & 7.2 & 9.2 & 90.8 \\
Position 18 & 13.2 & 21.6 & 588 & 8.8 & 11.3 & 99.4 & 8.7 & 10.5 & 116 \\
Position 19 & 10.9 & 12.9 & 74.5 & 9.8 & 12.2 & 82.5 & 7.1 & 8.6 & 146 \\
Position 20 & 7.2 & 8.8 & 52.4 & 8.9 & 11.2 & 101 & 6.8 & 8.8 & 54.2 \\
Position 21 & 11.1 & 13.8 & 83.6 & 8.8 & 11.2 & 98.7 & 6.0 & 7.5 & 107 \\
Position 22 & 8.7 & 10.7 & 63.3 & 9.5 & 11.8 & 101 & 7.1 & 9.2 & 161 \\
Position 23 & 9.1 & 11.2 & 72.4 & 10.2 & 12.7 & 146 & 8.9 & 10.4 & 89.8 \\
Position 24 & 7.8 & 10.0 & 72.7 & 9.5 & 12.6 & 144 & 7.0 & 8.6 & 270 \\
Position 25 & 8.6 & 10.9 & 165 & 10.9 & 13.7 & 107 & 5.9 & 8.3 & 314 \\
\hline
Mean & 9.2 & \color{red}11.8 & 126.4 & 9.3 & \color{red}11.8 & 149.0 & 7.1 & \color{red}9.3 & 179.8 \\
\cdashline{1-10}
Min & 7.1 & 8.8 & 52.4 & 6.8 & 8.6 & 56.9 & 5.5 & 7.2 & 54.2 \\
\cdashline{1-10}
Max & 13.2 & 21.6 & 588.0 & 11.5 & 15.0 & 1154.0 & 9.5 & 17.4 & 859.0 \\
\cdashline{1-10}
Standard Dev. & 1.4 & 2.4 & 107.8 & 1.2 & 1.6 & 212.0 & 1.0 & 2.0 & 155.6 \\
\hline
\end{tabular}
\label{tab:SurfaceRoughnessMirror}
\end{table}
}

Therefore, no direct conclusion can be drawn regarding the effectiveness of laser remelting based on the surface roughness data alone. Note that melt pool boundaries formed by AM can be seen on the non-remelted mirror topography in Figure \ref{fig:surfacetopology} \textit{(a)}, however the effect these boundaries have on surface roughness has been masked by the influence of the SPDT tool marks; melt pool boundaries have been reported to influence the surface morphology of additively manufactured AlSi10Mg components after machining and SPDT\cite{Wang25}.

\begin{figure}
    \centering
    \includegraphics[width=1\linewidth]{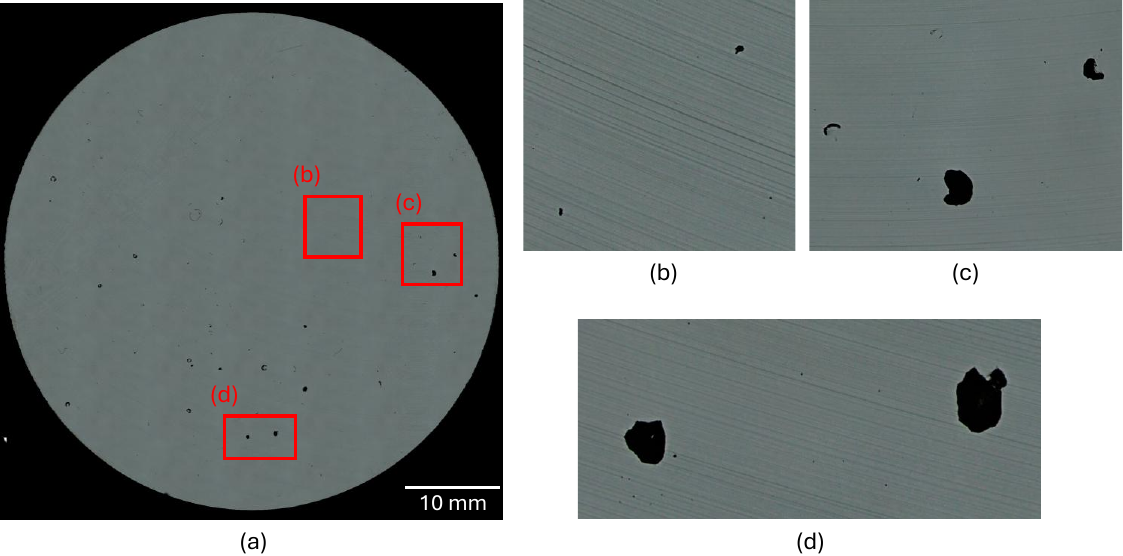}
    \caption{Examination of BF optical microscopy image on remelted mirror: \textit{a)} full surface, detail view locations annotated; \textit{b)} Small black spots, pore or oxide; \textit{c)} Keyhole pores; \textit{d)} Keyhole pores.}
    \label{fig:mirrorporesremelted}
\end{figure}

\begin{figure}
    \centering
    \includegraphics[width=1\linewidth]{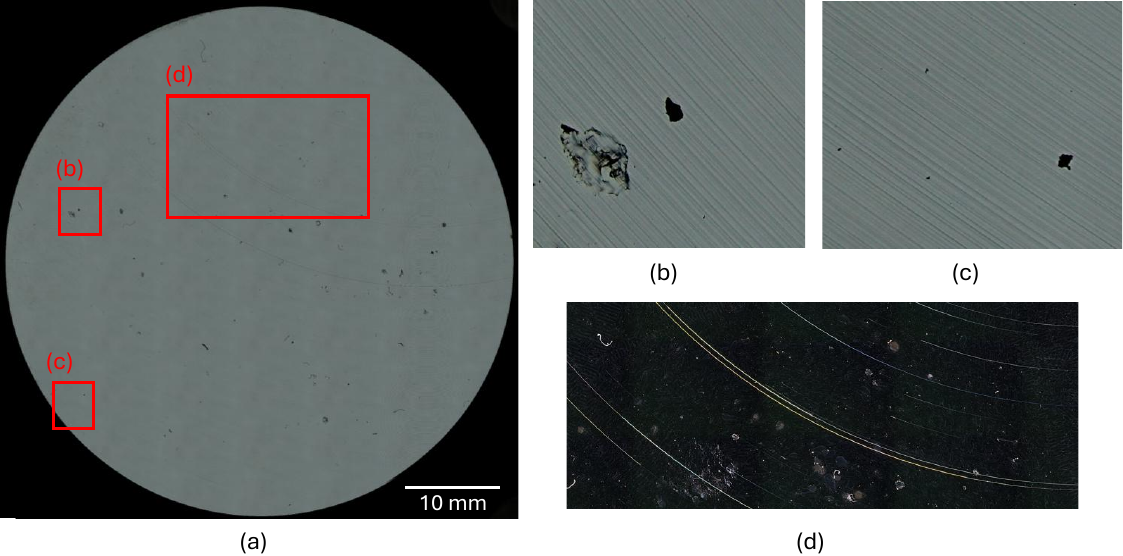}
    \caption{Examination of BF optical microscopy image on non-remelted mirror: \textit{a)} full surface, detail view locations annotated; \textit{b)} lack of fusion pore; \textit{c)} Small black spots, pore or oxide; \textit{d)} Large scratches on DF image.}
    \label{fig:nonremeltedpores}
\end{figure}

Examining the DF optical microscopy images in Figure \ref{fig:DFoptical} reveals a substantially larger amount of surface artefacts on the remelted and non-remelted mirrors than seen on the proof-of-concept cubes,  with numerous black spots distributed across the optical surface, some sufficiently large to be visible without magnification. Unlike the cube samples, where SEM, EDX, and FIB analysis confirmed the majority of black spots to be oxides, these larger defects observed on the optical surfaces have been identified as pores based on visual observation. Furthermore, two distinct pore morphologies can be identified: rounded pores consistent with keyhole porosity, and irregular shaped lack-of-fusion pores, are seen between the remelted and non-remelted mirrors in Figures \ref{fig:mirrorporesremelted} and \ref{fig:nonremeltedpores} respectively. The presence of both pore types indicates that multiple porosity formation mechanisms were active during the LPBF process within the optical surface region in both the remelted and non-remelted mirrors, suggesting further manufacturing parameter optimisation is required.

\begin{figure}
    \centering
    \includegraphics[width=1\linewidth]{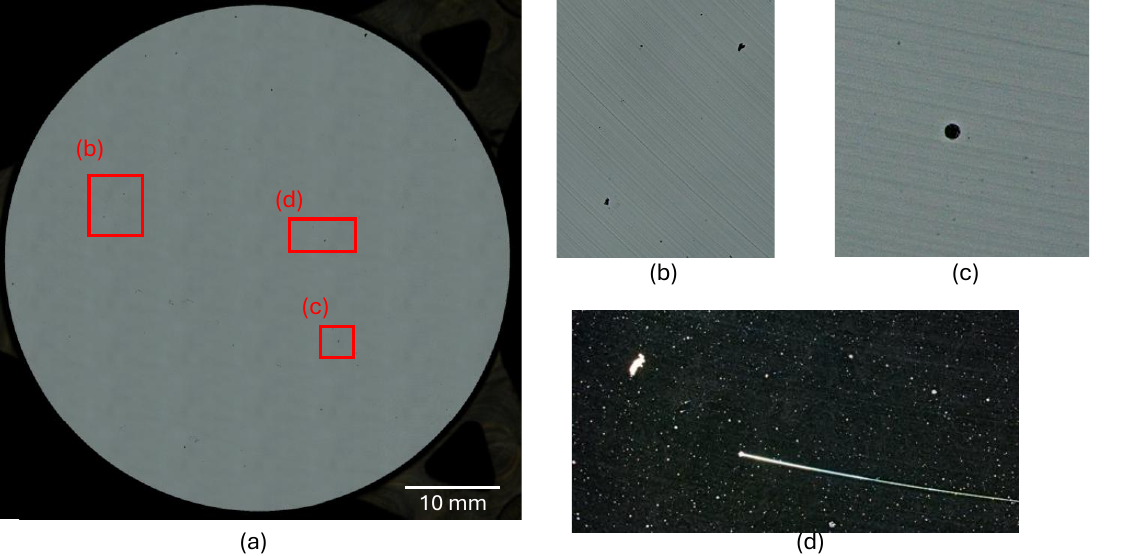}
    \caption{Examination of BF optical microscopy image on control mirror: \textit{a)} full surface, detail view locations annotated; \textit{b)} Small black spots, pore or oxide; \textit{c)} Small black spot, pore or oxide; \textit{d)} Large scratch on DF image.}
    \label{fig:controldefects}
\end{figure}

\begin{figure}
    \centering
    \includegraphics[width=1\linewidth]{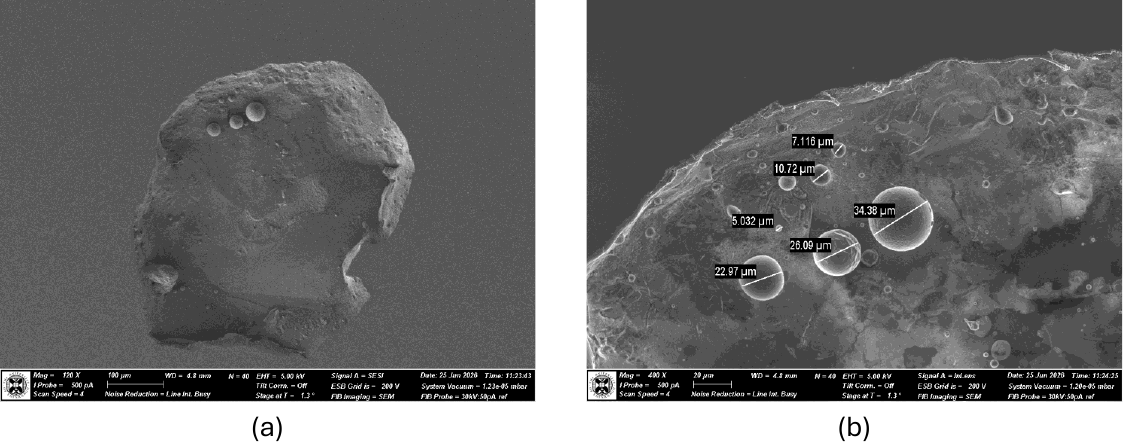}
    \caption{Lack-of-fusion pore sampled from the remelted optical surface.}
    \label{fig:SEMremelt}
\end{figure}

For comparison, the conventionally manufactured control mirror exhibited considerably fewer defects across the optical surface with a small number of black spots observed and a scratch initiated by one of these spots, as seen in Figure \ref{fig:controldefects}. This behaviour is consistent with the observations made during the proof-of-concept study, where embedded oxides were shown to act as scratch initiators during SPDT. Figure \ref{fig:SEMremelt} \textit{(a)} shows a lack-of-fusion pore sampled from the surface of the remelted mirror, confirmed by the measured unfused particles in Figure \ref{fig:SEMremelt} \textit{(b)}.

The results of this metrology indicate that the implementation of laser remelting on the first set of mirrors did not reproduce the exceptionally clean optical surface achieved on the remelted proof-of-concept cube samples. While the proof-of-concept study demonstrated that laser remelting could produce optical surfaces with no pores identified within the sampled regions and only a limited number of oxide-related defects, all AM mirrors contained a large population of visible defects and pores. Therefore, further optimisation of the LPBF process and SPDT post-processing is required before the benefits demonstrated on the cube samples can be translated to full-scale mirror components.

\section{Conclusion and future work}

This paper investigated the implementation and effectiveness of a laser remelting strategy that modifies the LPBF process to reduce porosity and improve the optical surface quality of AlSi10Mg AM mirrors. The strategy involves fully automated alternation between building and laser remelting throughout the manufacturing process. A preliminary proof-of-concept study that investigated remelted AM cubes with various manufacturing parameters demonstrated promising results: upon generating an optical surface on the cubes via SPDT, metrology results characterised one of the cubes with no pores detected in the sampled regions of its optical surface while matching the surface roughness of control cubes made via conventional manufacturing processes. This demonstrated significant potential for the laser remelting strategy, which can eliminate porosity in the bulk AM material, and allow the production of an optical surface via SPDT with surface roughness values comparable to conventionally manufactured aluminium mirrors.

Building on these findings, a first attempt at reproducing these results on a \diameter \SI{52}{\mm} sandwich mirror, designed for laser remelting while containing a lightweight lattice for 50\% mass reduction, was conducted. A control mirror was conventionally machined from an RSA 6061 billet, and a non-remelted mirror manufactured for comparison. After post-processing to generate an optical surface, the pore-free surfaces achieved during the proof-of-concept study was not reproduced on the AM mirror prototypes. Archimedes' principle tests predicted at least 1\% porosity throughout each AM mirror, while XCT analysis indicated 0.11\% in the sampled regions. Optical microscopy revealed a high density of visible surface defects on both the remelted and non-remelted mirrors, including porosity.

Surface roughness measurements for the first set of mirrors showed values higher than expected, with mean RMS of \SI{11.8}{\nm} for the non-remelted mirror, \SI{11.8}{\nm} for the remelted mirror, and \SI{9.3}{\nm} for the control mirror; based on the standard deviations, these values showed no clear statistical distinction between the mirrors. Furthermore, the surface topography of the sampled regions was dominated by the SPDT tool marks, rather than expected manufacturing defects.

Overall, while the proof-of-concept confirmed the viability of laser remelting to mitigate porosity on an optical surface, the results from the first set of prototypes demonstrate that further work is required to translate this result from simple geometries to complex mirrors. Further optimization of both the LPBF parameters and the SPDT post-processing workflow is ongoing, with a second set of mirrors with optimised manufacturing parameters currently being machined.

\section{DATA AVAILABILITY}
The dataset supporting this paper is openly available from eData the STFC Research Data repository at \url{https://doi.org/10.5286/edata/976}.

\acknowledgments 
The authors acknowledge the UKRI Future Leaders Fellowship ‘Printing the future of space telescopes’ under grant \# MR/T042230/1.
The authors also thank Liang Yang from Voxshell for access to ChopMesh, an add-on for nTop, the primary CAD software used in this paper.
Contributors of this project have also received funding from the European Union’s framework programme for research and innovation ‘Horizon Europe’ under the Marie Skłodowska-Curie grant agreement No 101119917 project METRAMAT.


\bibliography{report} 

@inproceedings{Lister24,
author = {Greg Lister and Rhys Tuck and Younes Chahid and Katherine Morris and Richard Kotlewski and Scott McPhee and Cyril Bourgenot and Ken Parkin and Mat Beardsley and Marta Civitani and Gabriele Vecchi and Carolyn Atkins},
title = {{Design, manufacture, and metrology of additively manufactured, metal, and ceramic lightweight circular mirror prototypes}},
volume = {13100},
booktitle = {Advances in Optical and Mechanical Technologies for Telescopes and Instrumentation VI},
editor = {Ram{\'o}n Navarro and Ralf Jedamzik},
organization = {International Society for Optics and Photonics},
publisher = {SPIE},
pages = {1310007},
year = {2024},
doi = {10.1117/12.3019053},
URL = {https://doi.org/10.1117/12.3019053}
}

@article{Ordnung24,
title = {Novel strategy for automated quality enhancement of up-facing inclined surfaces by incremental dual laser powder bed fusion},
journal = {Optics and Lasers in Engineering},
volume = {178},
pages = {108172},
year = {2024},
issn = {0143-8166},
doi = {https://doi.org/10.1016/j.optlaseng.2024.108172},
url = {https://www.sciencedirect.com/science/article/pii/S0143816624001519},
author = {Daniel Ordnung and Thibault Mertens and Jitka Metelkova and Brecht {Van Hooreweder}}
}

@inproceedings{Aziz25,
author = {Ilhan Aziz and Younes Chahid and Jennifer Keogh and James Carruthers and Katherine Morris and Joel Harman and Scott McPhee and Eilidh Fraser and Luca Millan and Cyril Bourgenot and Paul White and Spencer Davies and Franck P. Vidal and Wenjuan Sun and Mirko Sinico and Fraser Laidlaw and Wai Jue Tan and Arindam Majhi and Carolyn Atkins},
title = {{Additive manufacturing in aluminium of a primary mirror for a CubeSat application: manufacture, testing, and evaluation}},
volume = {13624},
booktitle = {Astronomical Optics: Design, Manufacture, and Test of Space and Ground Systems V},
editor = {Tony B. Hull and Daewook Kim and Pascal Hallibert},
organization = {International Society for Optics and Photonics},
publisher = {SPIE},
pages = {136241Y},
year = {2025},
doi = {10.1117/12.3063056},
URL = {https://doi.org/10.1117/12.3063056}
}

@inproceedings{Atkins19-1,
author = {Carolyn Atkins and William Brzozowski and Naomi Dobson and Maria Milanova and Stephen Todd and David Pearson and Cyril Bourgenot and David Brooks and Robert Snell and Wenjuan Sun and Peter Cooper and Simon G Alcock and Ioana-Theodora Nistea},
title = {{Additively manufactured mirrors for CubeSats}},
volume = {11116},
booktitle = {Astronomical Optics: Design, Manufacture, and Test of Space and Ground Systems II},
editor = {Tony B. Hull and Dae Wook Kim and Pascal Hallibert},
organization = {International Society for Optics and Photonics},
publisher = {SPIE},
pages = {1111616},
year = {2019},
doi = {10.1117/12.2528119},
URL = {https://doi.org/10.1117/12.2528119}
}

@inproceedings{Westsik23,
author = {Marcell Westsik and James T. Wells and Younes Chahid and Katherine Morris and Maria Milanova and Mat Beardsley and Michael Harris and Lucas Ward and Simon G. Alcock and Ioana-Theodora Nistea and Simone Cottarelli and Samuel Tammas-Williams and Carolyn Atkins},
title = {{From design to evaluation of an additively manufactured, lightweight, deployable mirror for Earth observation}},
volume = {12677},
booktitle = {Astronomical Optics: Design, Manufacture, and Test of Space and Ground Systems IV},
editor = {Tony B. Hull and Daewook Kim and Pascal Hallibert},
organization = {International Society for Optics and Photonics},
publisher = {SPIE},
pages = {1267704},
year = {2023},
doi = {10.1117/12.2677303},
URL = {https://doi.org/10.1117/12.2677303}
}

@article{Wang25,
title = {Effect of pre-heat treatment on the inherent feature and diamond machined surface morphologies of AlSi10Mg manufactured by laser powder bed fusion},
journal = {Materials Today Communications},
volume = {42},
pages = {111428},
year = {2025},
issn = {2352-4928},
doi = {https://doi.org/10.1016/j.mtcomm.2024.111428},
url = {https://www.sciencedirect.com/science/article/pii/S2352492824034111},
author = {Yifan Wang and Jun Yu and Pengfeng Sheng and Jingjing Xia and Di Zhu and Weichen Gu and Ruohui Xian and Xiaotian Liu and Dongfang Wang and Fei Han and Zhanshan Wang}
}

@inproceedings{Tan24,
author = {Songnian Tan and Yongsen Xu and Yefei Wang and Lei Shi},
title = {{Optical surface roughness improvement using hot isostatic pressing and coating of additively manufactured mirrors}},
volume = {13183},
booktitle = {International Conference on Optoelectronic Information and Functional Materials (OIFM 2024)},
editor = {Rachid Masrour and Xuemei Xu},
organization = {International Society for Optics and Photonics},
publisher = {SPIE},
pages = {131831W},
year = {2024},
doi = {10.1117/12.3034060},
URL = {https://doi.org/10.1117/12.3034060}
}

@article{Tan20,
author = {Songnian Tan and Yalin Ding and Yongsen Xu and Lei Shi},
title = {{Design and fabrication of additively manufactured aluminum mirrors}},
volume = {59},
journal = {Optical Engineering},
number = {1},
publisher = {SPIE},
pages = {013103},
year = {2020},
doi = {10.1117/1.OE.59.1.013103},
URL = {https://doi.org/10.1117/1.OE.59.1.013103}
}

@article{Bai20,
title = {Optical surface generation on additively manufactured AlSiMg0.75 alloys with ultrasonic vibration-assisted machining},
journal = {Journal of Materials Processing Technology},
volume = {280},
pages = {116597},
year = {2020},
issn = {0924-0136},
doi = {https://doi.org/10.1016/j.jmatprotec.2020.116597},
url = {https://www.sciencedirect.com/science/article/pii/S0924013620300108},
author = {Yuchao Bai and Zhuoqi Shi and Yan Jin Lee and Hao Wang}
}

@inproceedings{Atkins18,
author = {Carolyn Atkins and Charlotte Feldman and David Brooks and Stephen Watson and William Cochrane and Melanie Roulet and Emmanuel Hugot and Mat Beardsley and Michael Harris and Christopher Spindloe and Simon G. Alcock and Ioana-Theodora Nistea and Christian Morawe and Fran{\c{c}}ois Perrin},
title = {{Topological design of lightweight additively manufactured mirrors for space}},
volume = {10706},
booktitle = {Advances in Optical and Mechanical Technologies for Telescopes and Instrumentation III},
editor = {Ram{\'o}n Navarro and Roland Geyl},
organization = {International Society for Optics and Photonics},
publisher = {SPIE},
pages = {107060I},
year = {2018},
doi = {10.1117/12.2313353},
URL = {https://doi.org/10.1117/12.2313353}
}

@inproceedings{Snell22,
author = {Robert Snell and Carolyn Atkins and Hermine Schnetler and Younes Chahid and Mat Beardsley and Michael Harris and Chenxi Zhang and Richard Pears and Ben Thomas and Henry Saunders and Alexander Sloane and George Maddison and Iain Todd},
title = {{Towards understanding and eliminating defects in additively manufactured CubeSat mirrors}},
volume = {12188},
booktitle = {Advances in Optical and Mechanical Technologies for Telescopes and Instrumentation V},
editor = {Ram{\'o}n Navarro and Roland Geyl},
organization = {International Society for Optics and Photonics},
publisher = {SPIE},
pages = {121880V},
year = {2022},
doi = {10.1117/12.2629935},
URL = {https://doi.org/10.1117/12.2629935}
}

@Article{Liu22,
AUTHOR = {Liu, Chen and Xu, Kai and Zhang, Yongqi and Hu, Haifei and Tao, Xiaoping and Zhang, Zhiyu and Deng, Weijie and Zhang, Xuejun},
TITLE = {Design and Fabrication of Extremely Lightweight Truss-Structured Metal Mirrors},
JOURNAL = {Materials},
VOLUME = {15},
YEAR = {2022},
NUMBER = {13},
ARTICLE-NUMBER = {4562},
URL = {https://www.mdpi.com/1996-1944/15/13/4562},
PubMedID = {35806686},
ISSN = {1996-1944},
DOI = {10.3390/ma15134562}
}

@Article{Zhang21,
AUTHOR = {Zhang, Kai and Qu, Hemeng and Guan, Haijun and Zhang, Jizhen and Zhang, Xin and Xie, Xiaolin and Yan, Lei and Wang, Chao},
TITLE = {Design and Fabrication Technology of Metal Mirrors Based on Additive Manufacturing: A Review},
JOURNAL = {Applied Sciences},
VOLUME = {11},
YEAR = {2021},
NUMBER = {22},
ARTICLE-NUMBER = {10630},
URL = {https://www.mdpi.com/2076-3417/11/22/10630},
ISSN = {2076-3417},
DOI = {10.3390/app112210630}
}

@inproceedings{Cooper19,
Author = {Cooper, Peter A and Sun, Wenjuan and Atkins, Carolyn and Brown, Stephen B },
title = {Micro {XCT} measurements of defects in light-weighted mirrors for applications in space imaging},
volume = {Proc. of BINDT 2019},
year = {2019}
}

@article{kotadia21,
  volume = {46},
publisher = {Elsevier},
     doi = {10.1016/j.addma.2021.102155},
 journal = {Additive Manufacturing},
    year = {2021},
   title = {A review of laser powder bed fusion additive manufacturing of aluminium alloys : microstructure and properties},
   month = {October},
  author = {Kotadia, H. R. and Gibbons, Gregory John and Das, A. and Howes, P. D.},
     url = {http://dx.doi.org/10.1016/j.addma.2021.102155},
    issn = {2214-8604 }
}

@inproceedings{chahid24,
author = {Younes Chahid and Carolyn Atkins and Greg Lister and Rhys Tuck and Stephen Watson and Katherine Morris and David Isherwood and Jonathan Strachan and Joel Harman and Pearachad Chartsiriwattana and Deno Stelter and Werner Laun},
title = {{Additive manufacturing applications in astronomy: a review}},
volume = {13100},
booktitle = {Advances in Optical and Mechanical Technologies for Telescopes and Instrumentation VI},
editor = {Ram{\'o}n Navarro and Ralf Jedamzik},
organization = {International Society for Optics and Photonics},
publisher = {SPIE},
pages = {131001L},
year = {2024},
doi = {10.1117/12.3020182},
URL = {https://doi.org/10.1117/12.3020182}
}

@article{sun24,
title = {Influencing mechanisms of hot isostatic pressing on surface properties of additively manufactured AlSi10Mg alloy},
journal = {Journal of Materials Processing Technology},
volume = {329},
pages = {118426},
year = {2024},
issn = {0924-0136},
doi = {https://doi.org/10.1016/j.jmatprotec.2024.118426},
url = {https://www.sciencedirect.com/science/article/pii/S0924013624001444},
author = {Lijun Sun and Yulei Yang and Siyuan Li and Wencong Chen and Yichun Wang and Peng Yan and Yueqi Zhu and Weichao Wu and Bingliang Hu}
}

@article{kang25,
title = {Effect of hot isostatic pressing temperature on microstructures and characteristics of AlSi10Mg alloy fabricated by selective laser melting},
journal = {Journal of Materials Research and Technology},
volume = {37},
pages = {1443-1449},
year = {2025},
issn = {2238-7854},
doi = {https://doi.org/10.1016/j.jmrt.2025.06.122},
url = {https://www.sciencedirect.com/science/article/pii/S2238785425015443},
author = {Cheol Kang and Gun-Hee Kim and Won Rae Kim and Taeg Woo Lee and Min Ji Ham and Seung Jun Han and Seon-Jin Choi and Woo Jin Hwang and Young Jae Hwang and Hyun-Su Kang and Hyung Giun Kim}
}

@inproceedings{Atkins22,
author = {Carolyn Atkins and L. G. T. (Bart) van de Vorst and Andrew Conley and Szigfrid Farkas and Emmanuel Hugot and Gy{\"o}rgy Mező and Katherine Morris and M{\'e}lanie Roulet and Robert M. Snell and Fabio Tenegi-Sangin{\'e}s and Iain Todd and Afrodisio Vega-Moreno and Hermine Schnetler},
title = {{The OPTICON A2IM Cookbook: an introduction to additive manufacture for astronomy}},
volume = {12188},
booktitle = {Advances in Optical and Mechanical Technologies for Telescopes and Instrumentation V},
editor = {Ram{\'o}n Navarro and Roland Geyl},
organization = {International Society for Optics and Photonics},
publisher = {SPIE},
pages = {121880W},
year = {2022},
doi = {10.1117/12.2627244},
URL = {https://doi.org/10.1117/12.2627244}
}

@article{Katgerman04,
author = {L. Katgerman and Fred Dom},
title = {Rapidly solidified aluminium alloys by meltspinning},
journal = {Materials Science \& Engineering A: Structural Materials: Properties, Microstructure and Processing},
year = {2004},
volume = {375-377},
publisher = {Elsevier},
month = {jul},
url = {https://doi.org/10.1016/j.msea.2003.10.094},
pages = {1212--1216},
doi = {10.1016/j.msea.2003.10.094}
}

@inproceedings{Guido08,
author = {Guido P. H. Gubbels and Bart W. H. van Venrooy and Albert J. Bosch and Roger Senden},
title = {{Rapidly solidified aluminium for optical applications}},
volume = {7018},
booktitle = {Advanced Optical and Mechanical Technologies in Telescopes and Instrumentation},
editor = {Eli Atad-Ettedgui and Dietrich Lemke},
organization = {International Society for Optics and Photonics},
publisher = {SPIE},
pages = {70183A},
year = {2008},
doi = {10.1117/12.788766},
URL = {https://doi.org/10.1117/12.788766}
}

@article{Chahid24-pore,
author = {Chahid, Younes and Packer, C and Tawfik, Ahmed and Keen, J and Brewster, N and Beardsley, M and Morris, K and Bills, Paul and Blunt, Liam and Atkins, C and Tammas-Williams, Sam},
year = {2024},
month = {01},
pages = {},
title = {Development of a modular system to provide confidence in porosity analysis of additively manufactured components using x-ray computed tomography},
volume = {35},
journal = {Measurement Science and Technology},
doi = {10.1088/1361-6501/ad1670}
}

@online{nTopFDD,
  author       = {{nTop}},
  title        = {Field-Driven Design for Advanced Manufacturing, nTopology Inc., New York.},
  organization = {nTopology Inc.},
  address      = {New York, NY, USA},
  year         = {2026},
  url          = {https://www.ntop.com/field-driven-design/},
  urldate      = {2026-06-05},
}

@article{Du23,
title = {Pore defects in Laser Powder Bed Fusion: Formation mechanism, control method, and perspectives},
journal = {Journal of Alloys and Compounds},
volume = {944},
pages = {169215},
year = {2023},
issn = {0925-8388},
doi = {https://doi.org/10.1016/j.jallcom.2023.169215},
url = {https://www.sciencedirect.com/science/article/pii/S0925838823005182},
author = {Chuanbin Du and Yanhua Zhao and Jingchao Jiang and Qian Wang and Haijin Wang and Nan Li and Jie Sun}
}

@article{Atkins26,
doi = {10.1088/2631-8695/ae545f},
url = {https://doi.org/10.1088/2631-8695/ae545f},
year = {2026},
month = {mar},
publisher = {IOP Publishing},
volume = {8},
number = {7},
pages = {075403},
author = {Atkins, Carolyn and Westwood, Dominic and Aziz, Ilhan and Chahid, Younes and Laidlaw, Fraser H J and McPhee, Scott and Snell, Robert M and Tammas-Willams, Samuel},
title = {Mirror fabrication using metal additive manufacturing in AlSi10Mg: powder contamination and mitigation methods},
journal = {Engineering Research Express}
}

@article{TammasWilliams2017,
  author = {Tammas-Williams, S. and Withers, P. J. and Todd, I. and Prangnell, P. B.},
  title = {The Influence of Porosity on Fatigue Crack Initiation in Additively Manufactured Titanium Components},
  journal = {Scientific Reports},
  volume = {7},
  pages = {7308},
  year = {2017},
  doi = {10.1038/s41598-017-06504-5}
}

@inproceedings{Atkins2024,
author = {Carolyn Atkins and Younes Chahid and Gregory Lister and Rhys Tuck and Richard Kotlewski and Robert M. Snell and Elaine R. Livera and Mariam Faour and Iain Todd and Robert Deffley and James Shipley and Tom Walsh and Johannes G{\aa}rdstam and Cyril Bourgenot and Paul White and Spencer Davies and Samuel Tammas-Williams},
title = {{Targeting low micro-roughness for 3D printed aluminium mirrors using a hot isostatic press}},
volume = {13100},
booktitle = {Advances in Optical and Mechanical Technologies for Telescopes and Instrumentation VI},
editor = {Ram{\'o}n Navarro and Ralf Jedamzik},
organization = {International Society for Optics and Photonics},
publisher = {SPIE},
pages = {131003V},
year = {2024},
doi = {10.1117/12.3020239},
URL = {https://doi.org/10.1117/12.3020239}
}

@Article{Zhang22,
AUTHOR = {Zhang, Jizhen and Wang, Chao and Qu, Hemeng and Guan, Haijun and Wang, Ha and Zhang, Xin and Xie, Xiaolin and Wang, He and Zhang, Kai and Li, Lijun},
TITLE = {Design and Fabrication of an Additively Manufactured Aluminum Mirror with Compound Surfaces},
JOURNAL = {Materials},
VOLUME = {15},
YEAR = {2022},
NUMBER = {20},
ARTICLE-NUMBER = {7050},
URL = {https://www.mdpi.com/1996-1944/15/20/7050},
PubMedID = {36295122},
ISSN = {1996-1944},
DOI = {10.3390/ma15207050}
}
\bibliographystyle{spiebib} 

\end{document}